%% file: root.tex
\documentclass[journal]{IEEEtran}

\input{configuration.tex}
\input{header.tex}
\begin{document}
    \maketitle
        \input{abstract.tex}
    \input{paper.tex}

\input{appendices.tex}\input{acknowledgements.tex}
    \bibliographystyle{IEEEtran}
    \bibliography{bibliography}
\end{document}

%% file: configuration.tex
    \usepackage[utf8]{inputenc}
    \usepackage[english]{babel}
    \usepackage[T1]{fontenc}
    \usepackage{amsfonts}
    \usepackage{amsmath}
    \usepackage{mathtools}
    \usepackage{amssymb}
    \usepackage{gensymb}
    \usepackage{color}
    \usepackage{hhline}
    \usepackage{tikz} \usetikzlibrary{backgrounds}
    \usepackage{pgfplots} \pgfplotsset{compat=1.7}
    \usepackage{cite}
    \usepackage{filecontents}
    \usepackage{pgfplots, pgfplotstable}
    \usepgfplotslibrary{statistics}
    \usepackage{relsize}
    \usepackage[normalem]{ulem}
    \usepackage{subcaption}

    \newcommand{\iid}{\emph{i.i.d.}}
    \newcommand{\etal}{\emph{et al.}}
    \newcommand{\ie}{\emph{i.e.}}
    \newcommand{\eg}{\emph{e.g.}}

    \newcommand{\eqif}{\text{if }}
    \newcommand{\eqand}{\text{ and }}
    \newcommand{\eqotherwise}{\text{otherwise}}
    
    \newcommand{\figref}[1]{Fig.~\ref{#1}}
    
    \newcommand{\secref}[1]{Section~\ref{#1}}

    \newcommand{\defref}[1]{Definition~\ref{#1}}

    \newcommand{\pair}[2]{(#1, #2)}
    \newcommand{\unorderedPair}[2]{\{#1, #2\}}
    \newcommand{\intInterval}[2]{\{#1, \ldots, #2\}}
    \newcommand{\interval}[2]{[#1, #2]}
    \newcommand{\set}[1]{\{#1\}}
    \newcommand{\orderedSet}[1]{(#1)}
    \renewcommand{\~}[1]{\text{$\widetilde{#1}$}}
    \newcommand{\rec}[1]{\text{$\~{#1}$}}
    \newcommand{\spectral}[1]{\text{$\widehat{#1}$}}
    \newcommand{\mati}[2]{\text{$#1(#2)$}}
    \newcommand{\matij}[3]{\text{$#1(#2, #3)$}}
    \newcommand{\veci}[2]{\text{$#1(#2)$}}
    
    \newcommand{\G}{\text{$\mathcal{G}$}}
    \newcommand{\V}{\text{$\mathcal{V}$}}
    \newcommand{\E}{\text{$\mathcal{E}$}}
    \newcommand{\N}{\text{$N$}}
    \newcommand{\M}{\text{$M$}}
    \newcommand{\Q}{\text{$Q$}}
    
    \newcommand{\K}{\text{$K$}}
    \newcommand{\R}{\text{$R$}}
    \renewcommand{\P}{\text{$P$}}

    \renewcommand{\d}{\text{$d$}}
    \renewcommand{\u}{\text{$u$}}
    \renewcommand{\v}{\text{$v$}}
    
    \newcommand{\I}{\textbf{I}}
    \newcommand{\one}{\textbf{1}}
    \newcommand{\MU}{\textbf{M}}
    \newcommand{\W}{\textbf{W}}
    \newcommand{\D}{\textbf{D}}
    \newcommand{\A}{\textbf{A}}
    
    \newcommand{\T}{\textbf{T}}
    \newcommand{\NL}{\textbf{\L}}
    \newcommand{\NNL}{\textbf{L}}
    \newcommand{\x}{\textbf{x}}
    \newcommand{\y}{\textbf{y}}
    \newcommand{\e}{\textbf{e}}
    \renewcommand{\k}{\textbf{k}}

    \newcommand{\X}{\textbf{X}}
    \newcommand{\Y}{\textbf{Y}}
    \newcommand{\cov}{\text{$\boldsymbol{\Sigma}$}}
    \newcommand{\Th}{\text{$\boldsymbol{\Theta}$}}
    \newcommand{\eigval}{\text{$\boldsymbol{\Lambda}$}}
    
    \newcommand{\eigvec}{\text{$\boldsymbol{\mathcal{X}}$}}
    \newcommand{\el}{\text{$\lambda$}}
    \newcommand{\ev}{\text{$\boldsymbol{\chi}$}}
    \newcommand{\lNorm}[1]{\text{$\ell_{#1}$}}
    \newcommand{\LNorm}[2]{\text{$L_{#1,#2}$}}
    \newcommand{\norm}[2]{\text{$\|#1\|_{#2}$}}
    
    \newcommand{\REPRE}{\text{REPRE}}
    \newcommand{\MEPRE}{\text{MEPRE}}
    \newcommand{\recall}{\small{\text{recall}}}
    \newcommand{\precision}{\small{\text{precision}}}
    \newcommand{\fmeasure}{\small{\text{F-measure}}}
    
    \newtheorem{definition}{Definition}

    \DeclareMathOperator*{\trace}{\mathrm{Tr}}
    
    \DeclareMathOperator*{\diff}{\mathrm{diff}}
    
    \DeclareMathOperator*{\st}{\mathrm{s.t.}}
    \newcommand{\setR}{\text{$\mathbb{R}$}}

    \newcommand{\esp}[1]{\text{$\mathbb{E}\left[#1\right]$}}
    \newcommand{\espVar}[2]{\text{$\mathbb{E}_{#1}\left[#2\right]$}}
    \newcommand{\tr}[1]{\text{$#1^\top$}}

    \makeatletter\newcommand*{\rom}[1]{\expandafter\@slowromancap\romannumeral #1@}\makeatother
    
    \newcommand{\newText}{\color{black}}
    \newcommand{\stopNewText}{\color{black}}
    

%% file: header.tex
\title{Characterization and Inference of Graph Diffusion Processes from Observations of Stationary Signals}

\author{Bastien~Pasdeloup,
		Vincent~Gripon,
		Gr\'{e}goire~Mercier,
        Dominique~Pastor,
		and~Michael~G.~Rabbat%
	\thanks{This was supported by the European Research Council under the European Union's Seventh Framework Programme (FP7/2007-2013) / ERC grant agreement n\degree~290901, by the Labex CominLabs Neural Communications, and by the Natural Sciences and Engineering Research Council of Canada through grant RGPAS 429296-12.}
	\thanks{B.~Pasdeloup, V.~Gripon, G.~Mercier, and D.~Pastor are with UMR CNRS Lab-STICC, T\'{e}l\'{e}com Bretagne, 655 Avenue du Technopole, 29280, Plouzan\'{e}, France. Email: \{name.surname\}@telecom-bretagne.eu.}
	\thanks{M.G.~Rabbat is with the Department of Electrical and Computer Engineering, McGill University, 3480 University Street, Montr\'{e}al, H3A~0E9, Canada. Email: michael.rabbat@mcgill.ca.}%
}


%% file: abstract.tex
\begin{abstract}
    Many tools from the field of graph signal processing exploit knowledge of the underlying graph's structure (\eg, as encoded in the Laplacian matrix) to process signals on the graph.
    Therefore, in the case when no graph is available, graph signal processing tools cannot be used anymore.
    Researchers have proposed approaches to infer a graph topology from observations of signals on its nodes.
    Since the problem is ill-posed, these approaches make assumptions, such as smoothness of the signals on the graph, or sparsity priors.
    In this paper, we propose a characterization of the space of \emph{valid} graphs, in the sense that they can explain stationary signals.
    To simplify the exposition in this paper, we focus here on the case where signals were \iid{} at some point back in time and were observed after diffusion on a graph.
    We show that the set of graphs verifying this assumption has a strong connection with the eigenvectors of the covariance matrix, and forms a convex set.
    Along with a theoretical study in which these eigenvectors are assumed to be known, we consider the practical case when the observations are noisy, and experimentally observe how fast the set of valid graphs converges to the set obtained when the exact eigenvectors are known, as the number of observations grows.
    To illustrate how this characterization can be used for graph recovery, we present two methods for selecting a particular point in this set under chosen criteria, namely graph simplicity and sparsity.
    \newText
    Additionally, we introduce a measure to evaluate how much a graph is adapted to signals under a stationarity assumption.
    \stopNewText
    Finally, we evaluate how state-of-the-art methods relate to this framework through experiments on a dataset of temperatures.
\end{abstract}

%% file: paper.tex
    
    \section{Introduction}
    \label{introduction}
        
        In many applications, such as brain imaging \cite{Fallani2014} and hyperspectral imaging \cite{CampsValls2007}, it is convenient to model the relationships among the entries of the signals studied using a graph.
        Tools such as graph signal processing can then be used to help understand the studied signals, providing a spectral view of them.
        However, there are many cases where a graph structure is not readily available, making such tools not directly appliable.
        
        Graph topology inference from only the knowledge of signals observed on the vertices is a field that has received a lot of interest recently.
        Classical methods to obtain such a graph are generally based on estimators of the covariance matrix using tools such as covariance selection \cite{Dempster1972} or thresholding of the empirical covariance matrix \cite{Bickel2008}.
        More recent approaches make assumptions on the graph, and enforce properties such as sparsity of the graph and/or smoothness of the signals \cite{Lake2010, Dong2014, Kalofolias2016}.
        
        A common aspect of all these techniques is that they propose graph inference strategies that directly find a particular topology from the signals based on some priors.
        Rather than performing a direct graph inference, we explore an approach that proceeds in two steps.
        First, we characterize the matrices that may explain the relationships among signal entries.
        Then, we introduce criteria to select a matrix from this set.
        
        In this paper, we consider the case of stationary signals \cite{Girault2015b, Perraudin2016, Marques2016}.
        These signals are such that their covariance matrix has the same eigenvectors as the graph Fourier transform operator.
        To simplify the exposition in this paper, we focus here on the case of diffusion matrices, but the same ideas and methods could work for general observations of stationary signals on graphs.
        We assume that the signals were \iid{} at some point back in time.
        The relationships among entries of the signals were then introduced by a diffusion matrix applied a variable number of times on each signal.
        This matrix has non-null entries only when a corresponding edge exists in the underlying graph, and therefore is compliant with the underlying graph structure, modeling a diffusion process on it.
        Such matrices are referred to as \emph{graph shift operators} \cite{Sandryhaila2013, Sandryhaila2014}, examples of which are the adjacency matrix or the graph Laplacian.
        Under these settings, we address in this paper the following question: \emph{How can one characterize an adapted diffusion matrix from a set of observed signals?}
        
        \newText
        To answer this question, we choose to focus in this paper on a particular family of matrices to model the diffusion process for the signals.
        \stopNewText
        Similar results can be obtained with other graph shift operators by following the same development.
        
        We show that retrieving a diffusion matrix from signals can be done in two steps, by first characterizing the set of admissible candidate matrices, and then by introducing a selection criterion to encourage desirable properties such as sparsity or simplicity of the matrix to retrieve.
        This particular set of admissible matrices is defined by a set of linear inequality constraints.
        A consequence is that it is a convex polytope, in which one can select a point by defining a criterion over the set of admissible diffusion matrices and then maximizing or minimizing the criterion.
        
        We show that all candidate matrices share the same set of eigenvectors, namely those of the covariance matrix.
        Along with a theoretical study in which these eigenvectors are assumed to be known, we consider the practical case when only noisy observations of them are available, and observe the speed of convergence of the approximate set of solutions to the limit one, as the number of observed signals increases.

        Two criteria for selecting a particular point in this set are proposed.
        The first one aims to recover a graph that is simple, and the second one encourages sparsity of the solution in the sense of the \LNorm{1}{1} norm.
        \newText
        Additionally, we propose a method to obtain a diffusion matrix adapted to stationary signals, given a graph inferred with other methods based on other priors.
        \stopNewText
        
        This paper is organized as follows. First, \secref{definitions} introduces the problem addressed in this article, and presents the notions and vocabulary that are necessary for a full understanding of our work.
        Then, \secref{relatedWork} reviews the work that has been done in graph recovering from the observation of signals.
        \secref{characterization} studies the desired properties that characterize the admissible diffusion matrices, both in the ideal case and in the approximate one.
        \newText
        \secref{strategies} introduces methods to select an admissible diffusion matrix in the polytope based on a chosen criterion.
        \stopNewText
        Then, in \secref{experiments}, these methods are evaluated on synthetic data.
        Finally, in \secref{experimentsReal}, a dataset of temperatures in Brittany is studied, and additional experiments are performed to establish whether current state-of-the-art methods can be used to infer a valid diffusion matrix.
        
    %


    \section{Problem formulation}
    \label{definitions}
        
        \subsection{Definitions}
        
            We consider a set of \N{} random variables (vertices) of interest.
            Our objective is, given a set of \M{} realizations (signals) of these variables, to infer a diffusion matrix adapted to the underlying graph topology on which unknown \iid{} signals could have evolved to generate the given \M{} observations.
            
            \begin{definition}[Graph]
                \label{graph}
                A graph \G{} is a pair \pair{\V}{\E} in which $\V = \intInterval{1}{\N}$ is a set of \N{} vertices and $\E \subseteq \V \times \V$ is a set of edges.
                In the remainder of this document, we consider positively weighted undirected graphs.
                Therefore, we make no distinction between edges \pair{\u}{\v} and \pair{\v}{\u}.
                We denote such an edge by \unorderedPair{\u}{\v}.
                A convenient way to represent \G{} is through its adjacency matrix \W:
                $$\matij{\W}{\u}{\v} \triangleq \left\{
                  \begin{array}{cl}
                      \alpha_{\u\v} & \eqif \unorderedPair{\u}{\v} \in \E \\
                      0 & \eqotherwise
                  \end{array}
                  \right.; \alpha_{\u\v} \in \setR_+; \forall \u, \v \in \V \;.$$
            \end{definition}
            
            Graph shift operators are defined by Sandryhaila \etal{} \cite{Sandryhaila2013, Sandryhaila2014} as \emph{local operations that replace a signal value at each vertex of a graph with the linear combination of the signal values at the neighbors of that vertex}.
            The adjacency matrix is an example of graph shift operator, as its entries are non-null if and only if there exists a corresponding edge in \E.
            
            Graphs may have numerous properties that can be used as priors when inferring an unknown graph.
            In this paper, we are particularly interested in the sparsity of the graph, that measures the density of its edge, and in its simplicity.
            A graph is said to be \emph{simple} if no vertex is connected to itself, \ie, if the diagonal entries of the associated adjacency matrix are null.
            These two properties are often desired in application domains.
            \newText
            In the general case, we consider graphs that can have self-loops, \ie, non-null elements on the diagonal of \W.
            \stopNewText
            
            A signal on a graph can be seen as a vector that attaches a value to every vertex in \V.
            
            \begin{definition}[Signal]
                A signal \x{} on a graph \G{} of \N{} vertices is a function on \V.
                For convenience, signals on graphs are represented by vectors in $\setR^\N$, in which $\veci{\x}{i}$ is the signal component associated with the $i^\text{th}$ vertex of \V.
            \end{definition}
            
            A widely-considered matrix that allows the study of signals on a graph \G{} is the normalized Laplacian of \G.
            
            \begin{definition}[Normalized Laplacian]
                The normalized Laplacian \NL{} of a graph \G{} with adjacency matrix \W{} is a differential operator on \G, defined by $\NL \triangleq \I - \D^{-\frac{1}{2}} \W \D^{-\frac{1}{2}}$; where \D{} is the diagonal matrix of degrees of the vertices: $\matij{\D}{\u}{\u} \triangleq \sum_{\v \in \V} \matij{\W}{\u}{\v}; \forall \u \in \V$, and $\I$ is the identity matrix of size \N.
                Note that for \NL{} to be defined, \D{} must contain only non-null entries on its diagonal, which is the case when every vertex has at least one neighbor.
                \label{normalizedLaplacian}
            \end{definition}
            
            An interpretation of this matrix is obtained by considering the propagation of a signal \x{} on \G{} using \NL.
            By definition, $\NL \x = \I \x - (\D^{-\frac{1}{2}} \W \D^{-\frac{1}{2}}) \x$.
            Therefore, the normalized Laplacian models the variation of a signal \x{} when diffused through one step of a diffusion process represented by the graph shift operator $\T_\NL \triangleq \D^{-\frac{1}{2}} \W \D^{-\frac{1}{2}}$.
            \newText
            More generally, in this paper, we define diffusion matrices as follows:
            \stopNewText
            
            \newText
            \begin{definition}[Diffusion matrix]
                \label{diffusionMatrix}
                A diffusion matrix \T{} is a symmetric matrix such that
                \begin{itemize}
                    \item $\forall \u, \v \in \V : \matij{\T}{\u}{\v} \geq 0$;
                    \item $\el_1 = 1$;
                    \item $\forall i, \in \intInterval{2}{\N} : |\el_i| \leq 1$,
                \end{itemize}
                where $\el_i$ are the eigenvalues of $\T$, in descending order.
            \end{definition}
            \stopNewText
            
            \newText
            The idea behind these constraints is that we want to model a diffusion process by a matrix.
            Such process propagates signal components from vertex to vertex and consequently consists of positive entries indicating what quantity of signal is sent to the neighboring vertices.
            Enforcing all eigenvalues to have their modulus be at most $1$ imposes a scale factor, and has the interesting consequence to cause the series $\left(\T^i\x\right)_i$ to be bounded, for any signal $\x$.
            
            Note that by construction, the largest eigenvalue of $\T_\NL$ is $1$.
            In our experiments, we will use this particular matrix $\T_\NL$ to diffuse signals on the graph.
            Other popular matrices could be used instead to diffuse signals.
            For example, any polynome of $\T_\NL$ could be used \cite{Mei2015, Thanou2016, PetricMaretic2017}.
            \stopNewText

        \subsection{Graph Fourier transform}
        \label{gft}
            
            One of the cornerstones of signal processing on graphs is the analogy between the notion of frequency in classical signal processing and the eigenvalues of the Laplacian.
            The eigenvectors of the Laplacian of a binary ring graph correspond to the classical Fourier modes (see \eg{}, \cite{Shuman2013} for a detailed explanation).
            The lowest eigenvalues are analogous to low frequencies, while higher ones correspond to higher frequencies.
            Using this analogy, researchers have successfully been able to use graph signal processing techniques on non-ring graphs (\eg, \cite{Agaskar2013, Tremblay2014}).
            
            To be able to do so, the Laplacian matrix of the studied graph must be diagonalizable.
            Although it is a sufficient, but not necessary, condition for diagonalization, we only consider undirected graphs in this article (see \defref{graph}), for which the normalized Laplacian as defined in \defref{normalizedLaplacian} is symmetric.
            Note that there also exist definitions of the Laplacian matrix when the graphs are directed \cite{Chung2005}.
            
            To understand the link between diffusion of signals on the graph and the notion of \emph{smoothness} on the graph introduced in \secref{smoothness}, we need to introduce the graph Fourier transform \cite{Chung1997, Shuman2013}, that transports a signal \x{} defined on the graph into its spectral representation \spectral\x:
            
            \begin{definition}[Graph Fourier transform]
                Let $\eigval = \orderedSet{\el_1, \dots, \el_\N}$ be the set of eigenvalues of \NL, sorted by increasing value, and $\eigvec = \orderedSet{\ev_1, \dots, \ev_\N}$ be the matrix of associated eigenvectors.
                The graph Fourier transform of a signal \x{} is the projection of \x{} in the spectral basis defined by \eigvec: $\spectral\x \triangleq \tr{\eigvec} \x$.
                \spectral\x{} is a vector in $\setR^\N$, in which $\veci{\spectral\x}{i}$ is the spectral component associated with $\ev_i$.
            \end{definition}
            
            This operator allows the transportation of signals into a spectral representation defined by the graph.
            \newText
            Note that there exist other graph Fourier transform operators, based on the eigenvectors of the non-normalized Laplacian ${\bf L} \triangleq \D - \W$ or on those of the adjacency matrix \cite{Sandryhaila2013}.
            \stopNewText
            
            An important property of the normalized Laplacian states that the eigenvalues of \NL{} lie in the closed interval \interval{0}{2}, with the multiplicity of eigenvalue $0$ being equal to the number of connected components in the graph, and $2$ being an eigenvalue for bipartite graphs only \cite{Chung1997}.
            We obtain that the eigenvalues of $\T_\NL$ lie in the closed interval \interval{-1}{1}, with at least one of them being equal to $1$.
            Also, since $\T_\NL$ and \NL{} only differ by an identity, both matrices share the same set of eigenvectors.
            If the graph is connected then $\T_\NL$ has a single eigenvalue equal to $1$, being associated with a constant-sign eigenvector $\ev_1$:
            \begin{equation}
                \forall i \in \intInterval{1}{\N} : \veci{\ev_1}{i} = \sqrt{\frac{\matij{\D}{i}{i}}{\trace(\D)}} \;,
                \label{ev1}
            \end{equation}
            where \D{} is the matrix of degrees introduced in \defref{normalizedLaplacian}, and all other eigenvalues of $\T_\NL$ are strictly less than $1$.
            
            Therefore, diffusing a signal \x{} using $\T_\NL$ shrinks the spectral contribution of the eigenvectors of \NL{} associated with high eigenvalues more than those associated with lower ones.
            
            It is worth noting that since one of the eigenvalues of $\T_\NL$ is equal to $1$, then the contribution of the associated eigenvector $\ev_1$ does not change after diffusion.
            Therefore, after numerous diffusion steps, $\left(\veci{\spectral\x}{i}\right)_{i \in [2; \N]}$ become close to null and \x{} becomes stable on any non-bipartite graph.
            As a consequence, we consider in our experiments signals that are diffused a limited number of times.

        \subsection{Smoothness of signals on the graph}
        \label{smoothness}
        
            A commonly desired property for signals on graphs is smoothness.
            Informally, a signal is said to be smooth on the graph if it has similar entries where the corresponding vertices are adjacent in the graph.
            \newText
            In more details, given a diffusion matrix \T{} for a graph \G{}, smoothness of a signal \x{} can be measured via the following quantity:
            \begin{equation}
                S(\x) \triangleq \displaystyle\sum_{\unorderedPair{\u}{\v} \in \E} \matij{\T}{\u}{\v} \left( \veci{\x}{\u} - \veci{\x}{\v} \right)^2 \;.
                \label{eqSmoothness}
            \end{equation}
            \stopNewText
            
            From this equation, we can see that the lower $S(\x)$ is, the more regular are the entries of \x{} on the graph.
            \newText
            When using $\T_\NL$ as a diffusion matrix, signals that are \emph{low-frequency}, \ie, that mostly have a spectral contribution of the lower eigenvectors of the Laplacian, have a low value of $S(\x)$ and are then smooth on the graph.
            \stopNewText
            As mentioned above, diffusion of signals using $\T_\NL$ shrinks the contribution of eigenvectors of \NL{} associated with higher eigenvalues more than the contribution of the ones associated with lower eigenvalues. 
            Thence the property that diffused signals become low-frequency after some diffusion steps, and hence smooth on the graph.
            In addition to seeing diffusion as a link between graphs and signals naturally defined on them, this interesting property justifies the assumption, made in many papers, that signals should be smooth on a graph modeling their support \cite{Lake2010, Dong2014, Kalofolias2016}.

        \subsection{Stationarity of signals on the graph}
        \label{stationarity}
            
            Considering stationary signals is a very classical framework in traditional signal processing that facilitates the analysis of signals.
            Analogously, stationary processes on graphs have been recently defined to ease this analysis in the context of signal processing on graphs \cite{Girault2015b, Perraudin2016, Marques2016}.
            
            A random process on a graph is said to be (wide-sense) stationary if its first moment is constant over the vertex set and its covariance matrix is invariant with respect to the localization operator \cite{Marques2016}.
            In particular, white noise is stationary for any graph, and any number of applications of a graph shift operator on such noise leaves the process stationary.
            This implies that the covariance matrix of stationary signals shares the same eigenvectors as this particular operator (see \secref{diffusionBased} for details).

            Diffusion of signals is a particular case of stationary processing.
            The example we develop in this article when studying diffusion of signals through a matrix \T{} can be generalized to any stationary process and any graph shift operator, with only few adaptations.

        \subsection{Problem formulation}
        \label{problemFormulation}
            
            Using the previously introduced notions, we can formulate the problem we address in this paper as follows.
            Let $\X = \orderedSet{\x_1, \dots, \x_\M}, \x_i \in \setR^\N$, be a $\N \times \M$ matrix of \M{} observations, one per column.
            Let $\Y = \orderedSet{\y_1, \dots, \y_\M}, \y_i \in \setR^\N$, be a $\N \times \M$ unknown matrix of \M{} \iid{} signals; \ie, the entries $\matij{\Y}{i}{j}$ are zero-mean, independent random variables.
            Let $\k \in \setR_+^\M$ be an unknown vector of \M{} positive numbers,
            \newText
            corresponding to the number of times each signal is diffused before observation.
            \stopNewText
            
            Given $\X$, we aim to characterize the set of all diffusion matrices\footnote{Throughout this article we will denote recovered/estimated quantities using a tilde.}
            \rec\T{} such that there exist \Y{} and \k{} with:
            \begin{equation}
                \forall i \in \intInterval{1}{\M} : \x_i = \rec\T^{\veci{\k}{i}} \y_i
                \;.
            \end{equation}
            
            \newText
            This framework can be seen as a particular case of graph filters \cite{Sandryhaila2014}, containing only a monomial of the diffusion matrix.
            From a practical point of view, this corresponds to the setup where all signals are observed at a given time $t$, but have been initialized at various instants $t - \veci{\k}{i}$.
            More generally, all polynomials of the diffusion matrix share the same eigenvectors.
            The key underlying assumption in our work is that each observation is the result of passing white noise through a graph filter whose eigenvectors are the same as those of the normalized Laplacian.
            Consequently, our approach can be applied to any graph filter
            \begin{equation}
                \label{graphFilter}
                \forall i \in \intInterval{1}{\M} : \x_i = \sum_{j = 0}^\infty (\mathcal{K}_i)_j \rec\T^{j} \y_i
                \;,
            \end{equation}
            for $\M$ sequences $\mathcal{K}_1, \dots, \mathcal{K}_\M$.
            
            To summarize the following sections, we infer a diffusion matrix in two steps.
            First, we characterize the convex set of solutions using the method in \secref{characterization}.
            Then, we select a point from this set using some criteria on the matrix we want to infer.
            The strategies we propose are given in \secref{strategies}.
            \stopNewText
            
        %
        
    %
    

    \section{Related work}
    \label{relatedWork}
        
        While much effort has gone into inferring graphs from signals, the problem of characterizing the set of admissible graphs under diffusion priors is relatively new, and forms the core of our work.
        In this section we review related work on reconstructing graphs from the observation of diffused signals and make connections to the approach we consider.
        \newText
        Additional approaches exist but consider different signal models such as time series \cite{Mei2015, Shen2017}, band-limited signals \cite{Sardellitti2016} or combinations of localized functions \cite{Thanou2016, PetricMaretic2017}.
        \stopNewText
        
        \subsection{Estimation of the covariance matrix}
        \label{covarianceEstimation}
            
            As stated in the introduction, obtaining the eigenvectors of the covariance matrix is a cornerstone of our approach.
            They allow us to define a polytope limiting the set of matrices that can be used to model a diffusion process.
            
            Since the covariance matrix $\cov \triangleq \esp{\X \tr\X}$ is not obtainable in practical cases, a common approach involve estimating \cov{} using the sample covariance matrix \rec\cov:
            \begin{equation}
                \rec\cov \triangleq \frac{1}{\M-1} (\X - \MU) \tr{(\X - \MU)}\;,
                \label{sampleCovarianceMatrix}
            \end{equation}
            where $\MU(i,j) \triangleq \frac{1}{\M} \sum\limits_{k = 1}^{\M} \X(i, k)$ is an $\N \times \M$ matrix with each row containing the mean signal value for the associated vertex.
            An interesting property of this matrix is that its eigenvectors converge to those of the covariance matrix as the number of signals increases (see \secref{estimatedEigenvectors}).
            
            Other methods exist to infer a covariance matrix \cite{Cai2011, Cai2012, Wu2009, Xiao2012} and may be interesting to consider in place of the sample covariance matrix.
            \newText
            Methods for retrieving a sparse covariance matrix based on properties of its spectral norm are described in \cite{Cai2011} and \cite{Cai2012}.
            However, these works do not provide any information on the convergence rate of the eigenvectors of their solutions to the eigenvectors of the covariance matrix, as the number of signals increases.
            \stopNewText
            Similarly, \cite{Wu2009} and \cite{Xiao2012} retrieve covariance matrices that converge in operator norm or in distribution.
            An intensive study of covariance estimation methods could be interesting to find techniques that improve the convergence of eigenvectors.
            This paper focuses on the use of the sample covariance matrix.

        \subsection{Graphical lasso for graph inference}
        \label{graphicalLassoSection}
        
            A widely-used approach to provide a graph is the \emph{graphical lasso}~\cite{Friedman2008}, which recovers a sparse \emph{precision matrix} (\ie, inverse covariance matrix) \rec\Th{} under the assumption that the data are observations from a multivariate Gaussian distribution.
            The core of this method consists in solving the following problem,
            \begin{equation}
                \rec\Th = \underset{\Th \ge 0}{\operatorname{argmin}}\left(\trace(\rec\cov \Th) - \log \det(\Th) + \lambda \norm{\Th}{1}\right) \;,
                \label{graphicalLasso}
            \end{equation}
            where \rec\cov{} is the sample covariance matrix and $\lambda$ is a regularization parameter controlling sparsity.
            
            Numerous variations of this technique have been developed \cite{Rothman2008, Witten2011, Mazumder2012b, Tan2015}, and several applications have been using graphical lasso-based methods for inferring a sparse graph.
            Examples can be found for instance in the fields of neuroimaging \cite{Huang2009, Yang2015} or traffic modeling \cite{Sun2012}.
            
            What makes this method interesting, in addition to its fast convergence to a sparse solution, is a previous result from Dempster~\cite{Dempster1972}.
            In the \emph{covariance selection model}, Dempster proposes that the inverse covariance matrix should have numerous null off-diagonal entries.
            An additional result from Wermuth \cite{Wermuth1976} states that the non-null entries in the precision matrix correspond to existing edges in a graph that is representative of the studied data.
            
            Therefore, in our experiments, we evaluate whether considering the result of the graphical lasso as a graph makes it admissible or not to model a diffusion process.
            However, when considering \eqref{graphicalLasso}, we can see that the method does not impose any similarity between the eigenvectors of the covariance matrix and those of the inferred solution.
            For this reason, we do not expect this method to provide a solution that is admissible in our settings.
            
            Close to the graphical lasso, \cite{Pavez2016} and \cite{Egilmez2016} propose an algorithm to infer a precision matrix by adding generalized Laplacian constraints.
            While this allows for good recovery of the precision matrix, it proceeds in an iterative way by following a block descent algorithm that updates one row/column per iteration.
            As for the graphical lasso, it does not force the eigenvectors of the retrieved matrix to match those of the covariance matrix, and therefore does not match our stationarity assumption.
            \newText
            Interestingly, these methods could also be mentioned in the next section, dedicated to smoothness-based methods.
            In particular, \cite{Pavez2016} has pointed out that minimizing the quantity $\trace(\rec\cov \Th)$ promotes smoothness of the solution when $\Th$ is a graph Laplacian.
            Additionally, \cite{Rabbat2017} promotes sparsity of the inferred graph by applying a soft threshold to the precision matrix, and shows that the solution matches a smoothness assumption on signals.
            \stopNewText

        \subsection{Smoothness-based methods for graph inference}
        \label{smoothnessBased}
            
            Another approach to recover a graph is to assume that the signal components should be similar when the vertices on which they are defined are linked with a strong weight in \W, thus enforcing \emph{natural} signals on this graph to be low-frequency (smooth).
            Using the definition of smoothness of signals on a graph in \eqref{eqSmoothness}, we can see that the smaller $S(\x)$, the more regular the components of \x{} on the graph.
            
            A first work taking this approach has been proposed by Lake and Tenenbaum \cite{Lake2010}, in which they solve a convex optimization problem to recover a sparse graph from data to learn the structure best representing some concepts.
            More recently, Dong \etal{} \cite{Dong2014} have proposed a similar method that outperforms the one by Lake and Tenenbaum.
            \newText
            In order to find a graph Laplacian that minimizes $S$ in \eqref{eqSmoothness} for a set of signals, the authors propose an iterative algorithm that converges to a local solution, based on the resolution of the following problem:
            \begin{equation}
                \begin{array}{l}
                    \NNL^* = \arg\min\limits_{\NNL, \Y} \| \X - \Y \|_F^2 + \alpha \trace(\tr\Y \NNL \Y) + \beta \| \NNL \|_F^2 \\*
                    ~~\st{} \left\{\begin{array}{l}
                            \trace(\NNL) = \N \\
                            \NNL(i, j) = \NNL(j, i) \leq 0, i \neq j \\
                            \forall i \in \intInterval{1}{\N} : \sum_{j = 1}^\N \NNL(i, j) = 0
                        \end{array}\right. \;,
                    \label{dongOpti}
                \end{array}
            \end{equation}
            where $\NNL^*$ is the non-normalized Laplacian recovered, $\| \cdot \|_F$ is the Frobenius norm, \Y{} is a matrix in $\setR^{\N \times \M}$ that can be considered as a noiseless version of signals \X, and $\alpha$ and $\beta$ are regularization parameters controlling the distance between \X{} and \Y, and the sparsity of the solution.
            
            Kalofolias \cite{Kalofolias2016} proposes a unifying framework to improve the previous solutions of Lake and Tenenbaum, and Dong \etal, by proposing a better prior and reformulating the problem to optimize over entries of the (weighted) adjacency matrix rather than the Laplacian.
            An efficient implementation of his work is provided in the Graph Signal Processing Toolbox \cite{Perraudin2014}.
            \stopNewText
            His approach consists in rewriting the problem as an $\lNorm{1}$ minimization, that leads to naturally sparse solutions.
            Moreover, the author has shown that the method from Dong \etal{} could be encoded in his framework.
            
            
            \newText
            Graph inference with smoothness priors continues to receive a lot of interest.
            Recently, Chepuri \etal{} \cite{Chepuri2017} have proposed to infer a sparse graph on which signals are smooth, using an edge selection strategy.
            \stopNewText
            Finally, enforcing the smoothness property for signals defined on a graph has also been considered by Shivaswamy and Jebara \cite{Shivaswamy2010}, where a method is proposed to jointly learn the kernel of an SVM classifier and optimize the spectrum of the Laplacian to improve this classification.
            Contrary to our approach, Shivaswamy and Jebara~\cite{Shivaswamy2010} study a semi-supervised case, in which the spectrum of the Laplacian is learned based on a set of labeled examples.

        \subsection{Diffusion based methods for graph inference}
        \label{diffusionBased}
            
            Recently we proposed a third approach to recover a graph from diffused signals.
            In \cite{Pasdeloup2015}, we study a particular case of the problem we consider here, namely when \k{} is a known constant vector.
            Let \K{} denote the value in every entry of this vector.
            We show in \cite{Pasdeloup2015} that the covariance matrix of signals diffused \K{} times on the graph is equal to $\T^{2\K}$.
            This implies that we need to recover a particular root of the covariance matrix to obtain \T.
            In more details, if \Y{} is a matrix of mutually independent signals with independent entries, $\X = \T^\K \Y$, and \cov{} is the covariance matrix of \X, we have:
            \begin{equation}
                \cov = \esp{\X \tr\X} = \esp{\T^\K \Y \tr\Y \tr{{\T^\K}}} = \T^{2\K}
                \;,
            \end{equation}
            using the independence of \Y{} and the symmetry of \T.
            
            Thanks to \K{} being known, one could then retrieve a matrix \rec\T{} by diagonalizing \cov, taking the $2\K$-square root of the obtained eigenvalues, and solving a linear optimization problem to recover their missing signs.
            This reconstruction process was illustrated on synthetic cases, where a graph \G{} is generated, and \M{} \iid{} signals are diffused on it using the associated matrix $\T^\K$ to obtain \X \cite{Pasdeloup2015}.
            Experiments demonstrate that when using $\rec\cov = \T^{2\K}$ (which is the limit case when \M{} grows to infinity), we can successfully recover $\rec\T = \T$.

            However, this previous work has two principal limitations:
            \begin{enumerate}
                \item The number of diffusion steps \k{} is constant and known, which is a limiting assumption since in practical applications signals may be obtained after a variable, unknown number of diffusion steps.
                      In this work, we remove this assumption.
                      Taking the $2\K$-square root of the eigenvalues of \cov{} is therefore no longer possible.
                \item The number of observations \M{} is assumed to be infinite so that we have a perfect characterization of the eigenvectors of the covariance matrix.
                      We also address this assumption in this paper and show that the higher \M, the closer the recovered graph to the ground truth.
            \end{enumerate}
            
            Ongoing work by Segarra \etal{} \cite{Segarra2016b, Segarra2017}, initiated in \cite{Segarra2016a}, takes a similar direction.
            The authors propose a two-step approach, where they first retrieve the eigenvectors of a graph shift operator, and then infer the missing eigenvalues based on some criteria.
            They also study the case of stationary graph processes, for which the covariance matrix shares the same eigenbasis as the graph Fourier transform operator, and use this information to infer a graph based on additional criteria.
            
            \newText
            However, while the characterization of the set of solutions is identical to ours, our works differ in the matrix selection strategy.
            Segarra \etal{} \cite{Segarra2016b} focus on adjacency and Laplacian inference, while we aim at recovering a matrix modeling a diffusion process.
            Still, note that both of our works can be easily extended to any graph shift operator, by setting up the correct set of constraints.
            The authors of \cite{Segarra2016b} solve a slightly different problem, where they minimize the $\lNorm{1}$ norm of the inferred matrix under more constraints than ours, which describe a valid Laplacian matrix.
            In particular, they enforce the diagonal elements of the solution to be null, thus considering graphs that do not admit self-loops.
            In more details, they solve the following optimization problem:
            \begin{equation}
                {\bf S}^* = \displaystyle\arg\min_{{\bf S}, \el_1, \dots, \el_\N}~\| {\bf S} \|_1
                ~~\st{} \left\{\begin{array}{l}
                    {\bf S} = \sum_{i = 1}^\N \el_i \ev_i \ev_i^\top \\
                    {\bf S} \in \mathcal{S}
                \end{array}\right. \;,
                \label{segarraOpt}
            \end{equation}
            where ${\bf S}^*$ is the inferred graph shift operator, $\mathcal{S}$ is the set of admissible solutions delimited by their constraints, and $\ev_1, \dots, \ev_\N$ are the eigenvectors of the covariance matrix.
            Contrary to their approach, we aim at inferring a matrix that can be simple (see \secref{simpleStrategy}) or sparse (see \secref{sparseStrategy}), rather than selecting a sparse matrix from the set of simple matrices.
            Among other differences, we propose in \secref{regularizationStrategy} a method to approximate the solution of any graph inference strategy to make it match our stationary assumption on signals.
            Our work also explores how the polytope of solutions can be used to evaluate which graph, among a set of given graphs, is the most adapted to given signals.
            \stopNewText

        \subsection{Other related work}
        \label{otherWork}
            
            Shahrampour and Preciado \cite{Shahrampour2013, Shahrampour2015} study the context of network inference from stimulation of its vertices with noise.
            However, their method implies a series of \emph{node knockout} operations that need to individually intervene on the vertices.
            
            Also, we note that there exist methods that aim to recover a graph from the knowledge of its Laplacian spectrum \cite{Ipsen2002}.
            However, we do not assume that such information is available.
            
            \newText
            Finally, a recent work by Shafipour \etal{} \cite{Shafipour2017} has started to explore the problem of graph inference from non-stationary graph signals, which is a direct continuation of the work presented in this article and of the work by Segarra \etal.
            \stopNewText
        
        %
        
    %


    \section{Characterization of the set of admissible diffusion matrices}
    \label{characterization}
        
        \newText
        In this section, we show that the set of diffusion matrices verifying the properties in \defref{diffusionMatrix} is a convex polytope delimited by linear constraints depending on the eigenvectors of the covariance matrix of signals diffused on the graph.
        Then, we study the impact of a limited number of observations on the deformation of this polytope, due to imprecisions in the obtention of these eigenvectors.
        \stopNewText
        
        \newText
        \subsection{Characterization of the polytope of solutions}
        \label{admissibleMatrices}
        \stopNewText
            
            In the asymptotic case when \M{} is infinite, the covariance matrix \cov{} of the given signals \X{} is equal to a (fixed) power \K{} of the diffusion matrix.
            Thus, under these asymptotic settings, $\eigvec$ can be obtained using Principal Component Analysis on \X{} \cite{Pearson1901}.
            In the more global case when \k{} is a vector, the covariance matrix of the signals is a linear combination of multiple powers of \T, and has therefore the same set of eigenvectors, since all powers of a matrix share the same eigenvectors.
            \newText
            This is also the case when considering graph filters as in \eqref{graphFilter}.
            \stopNewText

            In more details, if we consider signals $\x_i = \T^\veci{\k}{i} \y_i$, we have the following development.
            We denote by \mati{\X}{i} the signal at the $i^\text{th}$ column of \X, and drop the constant factor and signals mean from \eqref{sampleCovarianceMatrix} for readability:
            \begin{equation}
                \begingroup
                    \renewcommand*{\arraystretch}{1.9}
                    \begin{array}{rcl}
                        \rec\cov & = & \displaystyle\sum_{i = 1}^\M \mati{\X}{i} \tr{\mati{\X}{i}} \\
                                 & = & \displaystyle\sum_{k \in \k} \displaystyle\sum_{i \st \atop \veci{\k}{i} = k} \T^k \mati{\Y}{i} \tr{\mati{\Y}{i}} \tr{{\T^k}} \\
                                 & = & \displaystyle\sum_{k \in \k} \T^k \left( \displaystyle\sum_{i \st \atop \veci{\k}{i} = k} \mati{\Y}{i} \tr{\mati{\Y}{i}} \right) \tr{{\T^k}} \\
                        \cov & = & \displaystyle\sum_{k \in \k} \T^{k} \espVar{\Y}{\displaystyle\sum_{i \st \atop \veci{\k}{i} = k} \mati{\Y}{i} \tr{\mati{\Y}{i}}} \tr{{\T^k}} \\
                             & = & \displaystyle\sum_{k \in \k} \left| \left\{i, \veci{\k}{i} = k\right\} \right| \T^{2k} \;,
                    \end{array}
                \endgroup
            \end{equation}
            which is a linear combination of various powers of \T, all having the same eigenvectors $\eigvec$.
            
            \newText
            Let us first consider the limit case when the eigenvectors \eigvec{} of \cov{} are available.
            \stopNewText
            Given the remarks in \secref{gft}, to recover an acceptable diffusion matrix \rec\T, we must find eigenvalues $\rec\eigval = \orderedSet{\rec{\el_1}, \dots, \rec{\el_\N}}$ such that:
            \begin{itemize}
                \item $\forall i, j \in \intInterval{1}{\N}; j \geq i : \matij{\rec\T}{i}{j} \geq 0$;
                \item Let $\ev_1$ be the constant-sign eigenvector in \eigvec: $\rec{\el_1} = 1$;
                \item $\forall i \in \intInterval{1}{\N} : \rec{\el_i} \in \interval{-1}{1}$.
            \end{itemize}
            
            Note that these constraints are driven by the will to recover a diffusion matrix as defined in \defref{diffusionMatrix}.
            If we were considering the case of other graph shift operators, these constraints would be different and would yield the definition of different constraints in \eqref{inequationsPositivity}.
            As an example, diffusion using a Laplacian matrix would imply the definition of constraints that enforce the diagonal entries to be positive and off-diagonal ones to be negative (see \cite{Segarra2016b}).
            \newText
            Similarly, aiming to recover the diffusion matrix $\T_\NL \triangleq \D^{-\frac{1}{2}} \W \D^{-\frac{1}{2}}$ associated with the normalized Laplacian would imply additional constraints.
            \stopNewText
            
            To illustrate how these properties translate into a set of admissible diffusion matrices, let us consider the randomly generated $3 \times 3$ symmetric adjacency matrix
            $\W = \begin{psmallmatrix}
                0.417~ & 0.302~ & 0.186 \\
                0.302~ & 0.147~ & 0.346 \\
                0.186~ & 0.346~ & 0.397 \\
            \end{psmallmatrix}$.
            We compute its associated matrix $\T_\NL$ and corresponding eigenvectors $\eigvec$.
            \newText
            This simulates a perfect retrieval of the eigenvectors of the covariance matrix of signals diffused by $\T_\NL$ on the graph.
            \stopNewText
            For all pairs $\pair{\rec{\el_2}}{\rec{\el_3}} \in \interval{-1}{1} \times \interval{-1}{1}$ (using a step of $10^{-2}$), \figref{fig1b} depicts those that allow the reconstruction of a diffusion matrix.
    
            \newsavebox{\matrixInFig} \savebox{\matrixInFig}{$\begin{psmallmatrix} 1 & 0 & 0 \\ 0 & \rec{\el_2} & 0 \\ 0 & 0 & \rec{\el_3} \end{psmallmatrix}$}
            \begin{figure}
                \centering
                \includegraphics[width=0.71\linewidth]{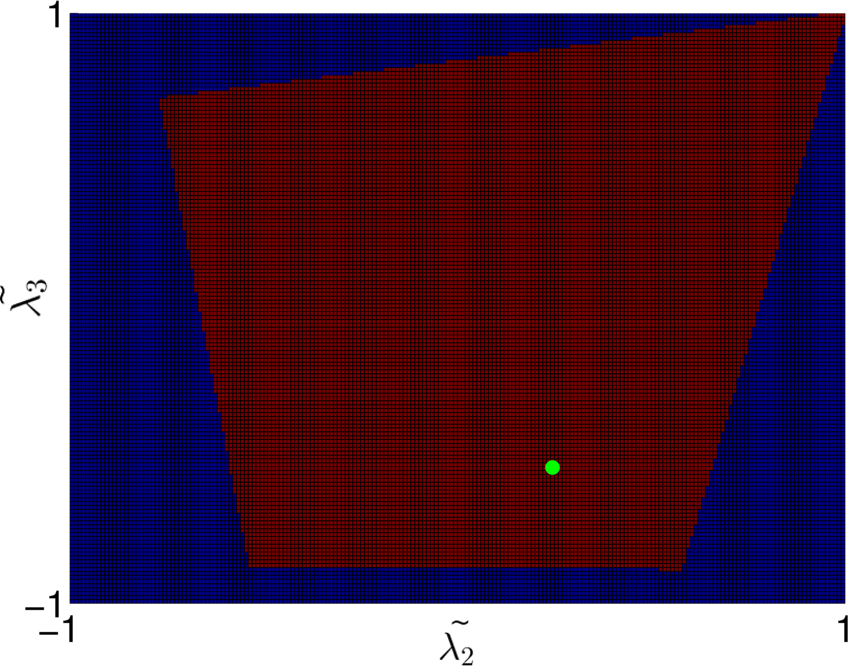}
                \caption
                {
                    All pairs $\pair{\rec{\el_2}}{\rec{\el_3}} \in \interval{-1}{1} \times \interval{-1}{1}$ (using a step of $10^{-2}$) for which $\rec\T = \eigvec \usebox{\matrixInFig} \tr{\eigvec}$ is an admissible diffusion matrix (in red).
                    The exact eigenvalues of the matrix $\T_\NL$ associated with \W{} are located using a green dot.
                }
                \label{fig1b}
            \end{figure}
            
            As we can see, the set of admissible matrices is convex (but non-strictly convex), delimited by affine equations.
            To characterize these equations, let us consider any entry of index $\pair{i}{j}$ in the upper triangular part of the matrix \rec\T{} we want to recover.
            Since \eigvec{} is assumed to be known, by developing the matrix product $\rec\T \triangleq \eigvec \begin{psmallmatrix}\rec{\el_1} & 0 & 0 \\ 0 & \dots & 0 \\ 0 & 0 & \rec{\el_\N}\end{psmallmatrix} \tr\eigvec$, we can write every entry \matij{\rec\T}{i}{j} as a linear combination of variables $\rec{\el_1}, \dots, \rec{\el_N}$ by developing the scalars in \eigvec.
            Let $\alpha_{ij1}, \dots, \alpha_{ij\N}$ be the factors associated with $\el_1, \dots, \el_N$ for the equation associated with the entry $\matij{\rec\T}{i}{j}$, \ie:
            \begin{equation}
                \matij{\rec\T}{i}{j} = \alpha_{ij1} \rec{\el_1} + \dots + \alpha_{ij\N} \rec{\el_\N} \;.
            \end{equation}
            
            As an example, let us consider a $3\times3$ matrix \T{} with known eigenvectors \eigvec.
            Using the decomposition of \T, we can write $\matij{\T}{2}{3}$ as follows:
            \begin{equation}
                \resizebox{\linewidth}{!}{
                    $\matij{\T}{2}{3} = \underbrace{\matij{\eigvec}{2}{1} \matij{\eigvec}{3}{1}}_{\mathlarger{\mathlarger{\alpha_{231}}}} \el_1 + \underbrace{\matij{\eigvec}{2}{2} \matij{\eigvec}{3}{2}}_{\mathlarger{\mathlarger{\alpha_{232}}}} \el_2 + \underbrace{\matij{\eigvec}{2}{3} \matij{\eigvec}{3}{3}}_{\mathlarger{\mathlarger{\alpha_{233}}}} \el_3 \;.$
                }
            \end{equation}
            
            Enforcing the value of all entries of the matrix to be positive thus defines the following set of $\frac{\N(\N+1)}{2}$ inequalities:
            \begin{equation}
                \forall i, j \in \intInterval{1}{\N} ; j \geq i : \alpha_{ij1} \rec{\el_1} + \dots + \alpha_{ij\N} \rec{\el_\N} \geq 0 \;,
                \label{inequationsPositivity}
            \end{equation}
            where $j \geq i$ comes from the symmetry property.
            
            \newText
            Our problem of recovering the correct set of eigenvalues to reconstruct the diffusion matrix thus becomes a problem of selecting a vector of dimension $\N-1$ (since one eigenvalue is equal to $1$ due to the imposed scale) in the convex polytope delimited by \eqref{inequationsPositivity}.
            \stopNewText
            Since the number of possible solutions is infinite, it is an ill-posed problem.
            To cope with this issue, one then needs to incorporate additional information or a selection criterion to enforce desired properties on the reconstructed matrix.
            To illustrate the selection of a point in the polytope, \secref{strategies} presents strategies based on different criteria, namely sparsity and simplicity.
            
            \newText
            Note that the polytope is in most cases not a singleton.
            The covariance matrix belongs to the set of admissible matrices, along with all its powers, which are different if the covariance matrix is not the identity.
            When additional constraints delimit the polytope, for example when enforcing the matrices to be simple, there exist situations when the solution is unique \cite{Segarra2016b}.
            \stopNewText

        \newText
        \subsection{Impact of the use of the sample covariance matrix on the polytope definition}
        \label{estimatedEigenvectors}
        \stopNewText
            
            The results discussed above use the eigenvectors \eigvec{} of the limit covariance matrix \cov{} as \M{} tends to infinity, directly obtained by diagonalizing the diffusion matrix of a ground truth graph under study.
            Next, we study the impact of the use of the sample covariance matrix \rec\cov{} of controlled signals respecting our assumptions on the estimation of these eigenvectors.
            
            To understand the impact of using \rec\cov{} instead of \cov{}, let us again consider an example $3 \times 3$ matrix and the associated polytope, as we did in \secref{admissibleMatrices}.
            We generate a random graph with $\N = 3$ vertices by drawing the entries of its adjacency matrix uniformly, and compute its matrix $\T_\NL$.
            Using $\T_\NL$, we diffuse \M{} \iid{} signals (entries are drawn uniformly) a variable number of times (chosen uniformly in the interval \interval{2}{5}) to obtain \X.
            Then, we compute the sample covariance matrix of \X, \rec\cov, and its matrix of eigenvectors \rec\eigvec.
            
\begin{figure*}
                \centering
                \begin{subfigure}{0.32\textwidth}
                    \includegraphics[width=\linewidth]{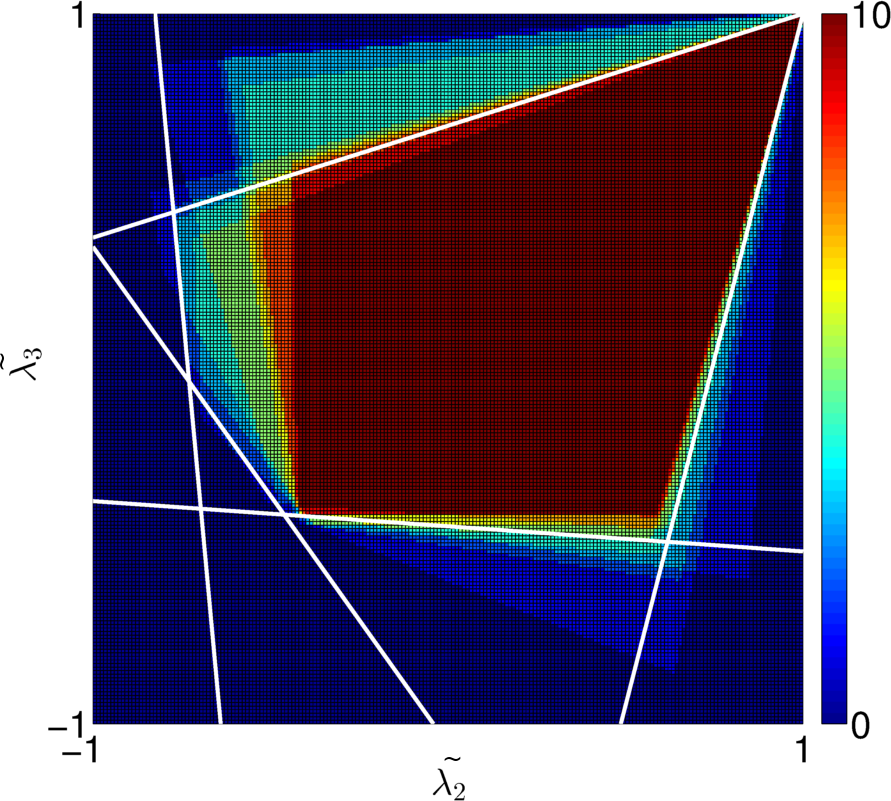}
                    \caption{}
                \end{subfigure}
                \begin{subfigure}{0.32\textwidth}
                    \includegraphics[width=\linewidth]{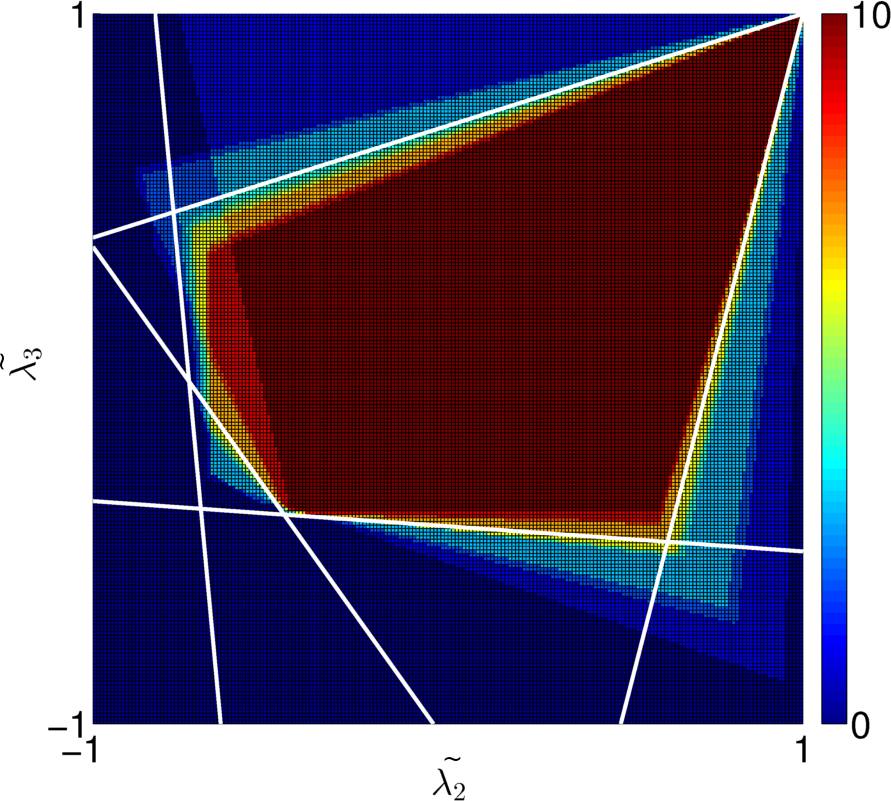}
                    \caption{}
                \end{subfigure}
                \begin{subfigure}{0.32\textwidth}
                    \includegraphics[width=\linewidth]{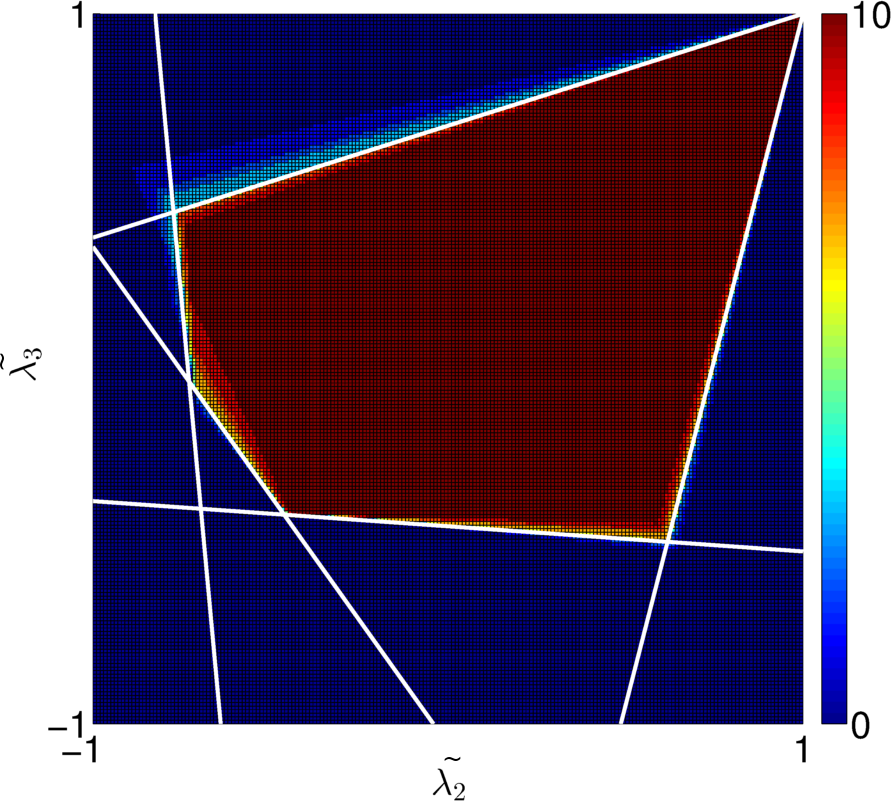}
                    \caption{}
                \end{subfigure}
                \caption
                {
                    Histogram representing the number of times a pair $\pair{\rec{\el_2}}{\rec{\el_3}}$ is valid, in the sense of the criteria in \secref{admissibleMatrices}, when used jointly with \rec\eigvec{} to recover a diffusion matrix.
                    The ground truth polytope is represented by the inequality constraints in white.
                    Results obtained for $10$ instances of \X{} on the same graph, for $\M = 10$ (a), $\M = 100$ (b) and $\M = 1000$ (c).
                }
                \label{polytopeNoise}
            \end{figure*}

            \figref{polytopeNoise} depicts in white the ground truth polytope (\ie, the one delimited by equations \eqref{inequationsPositivity} using the eigenvectors of \cov) and the polytopes associated with $10$ different sample covariance matrices \rec\cov, obtained from different realizations of \X.
            From each of the eigenvector sets of these empirical covariance matrices, we determine the pairs $\pair{\rec{\el_2}}{\rec{\el_3}}$ that satisfy the criteria in \secref{admissibleMatrices}.
            Then, we plot a histogram of the number of occurrences of these valid pairs.
            
            As we can see, the recovered polytope more accurately reflects the true one as \M{} increases.
            This coincides with the fact that the empirical covariance matrix converges to the real one as \M{} tends to infinity.
            
            In more details, we are interested in the convergence of the eigenvectors of the empirical covariance matrix $\rec\eigvec = \set{\rec{\ev_1}, \dots, {\rec\ev_\N}}$ to those of the actual covariance matrix \eigvec.
            Asymptotic results on this convergence are provided by Anderson \cite{Anderson1963}, which extends earlier related results by Girshick \cite{Girshick1939} and Lawley \cite{Lawley1956}.
            Let $\e_i \triangleq \tr\eigvec \rec{\ev_i}$ be the vector of cosine similarities between $\rec{\ev_i}$ and all eigenvectors of the actual covariance matrix.
            Anderson \cite{Anderson1963} states that, as the number of observations tends to infinity, entries in $\e_i$ have a Gaussian distribution with a known variance.
            In particular, when all eigenvalues are distinct, the inner product between the $i^\text{th}$ (for all $i$) eigenvector of the covariance matrix, $\ev_i$, and the $j^\text{th}$ (for all $j$) eigenvector of its estimate, $\rec{\ev_j}$, is asymptotically Gaussian with zero mean and variance
            \begin{equation}
                \frac{\el_i \rec{\el_j}}{(\M - 1) (\el_i - \rec{\el_j})^2}\;, \el_i \neq \el_j\;,
                \label{vectorsConvergence}
            \end{equation}
            where $\el_i$ is the eigenvalue associated with $\ev_i$, and $\rec{\el_j}$ is the eigenvalue associated with $\rec{\ev_j}$.
            As a consequence, the variance decreases like $\frac{1}{\M}$, and it also depends on the squared difference between $\el_i$ and $\rec{\el_j}$.
            Additionally, \cite{Anderson1963} shows that the maximum likelihood estimate $\rec{\el_i}$ of $\el_i$ (for all $i$) is
            \begin{equation}
                \rec{\el_i} = \frac{1}{\Q_i} \frac{\M - 1}{\M} \displaystyle\sum_{j \in \mathcal{L}_i} \el_j \;,
                \label{valuesConvergence}
            \end{equation}
            where $\Q_i$ is the multiplicity of eigenvalue $\el_i$, and $\mathcal{L}_i$ is the set of integers \set{$\Q_1 + \dots + \Q_{i-1} + 1, \dots, \Q_1 + \dots + \Q_i$}, containing all indices of equal eigenvalues.
            In the simple case when all eigenvalues are distinct, \eqref{valuesConvergence} simplifies to
            \begin{equation}
                \rec{\el_i} = \frac{\M - 1}{\M} \el_i \;.
                \label{valuesConvergence2}
            \end{equation}
            The eigenvalues of the empirical covariance matrix thus converge to those of the actual covariance matrix as \M{} increases.
            As \M{} tends to infinity, $\e_i$ thus tends to the $i^\text{th}$ canonical vector, indicating collinearity between $\rec{\ev_i}$ and $\ev_i$.
            Additionally, \cite{Anderson1963} provides a similar result for the more general case when eigenvalues may be repeated.
            
            \newText
            As we can see from \eqref{vectorsConvergence}, the convergence of the eigenvectors of the sample covariance matrix to the eigenvectors of the true covariance matrix is impacted by the eigenvalues of the matrix used to diffuse the signals.
            We know that diffusing a signal \K{} times using \T{} is equivalent to $\T^\K \x$.
            When rewriting this equation in the spectral basis, we obtain $(\eigvec \eigval^K \tr\eigvec) \x$, where $\eigvec$ and $\eigval$ are the eigenvectors and eigenvalues of \T.
            As we can see, the power distributes on the eigenvalues, and due to their location in the interval $]-1, 1]$ (with the noticeable exception of bipartite graphs), the term $(\el_i - \rec{\el_j})^2$ in \eqref{vectorsConvergence} gets smaller, and $\M$ must grow to achieve the same precision.
            
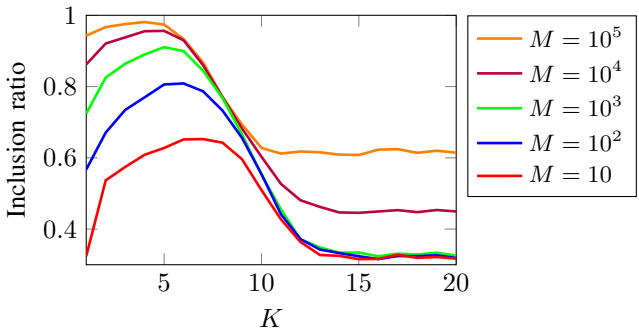
\begin{figure}
                \centering
                \input{inPolytopeVaryingMK.tex}
                \caption
                {
                    Ratio of cases when the eigenvalues of the ground truth matrix belong to the approximate polytope, as a function of the number of diffusions $\K$, for various quantities of signals $\M$.
                    Tests were performed for $10^4$ occurrences of random adjacency matrices of $\N = 10$ vertices.
                }
                \label{convergenceOfEigenvectorsKM}
            \end{figure}

            To illustrate the impact of the number of diffusions on the convergence of the polytope, let us consider the following experiment.
            We generate $10^4$ occurrences of random adjacency matrices of $\N = 10$ vertices, by drawing their entries uniformly in \interval{0}{1}, and enforcing symmetry.
            Then, for each adjacency matrix, we compute the associated matrix $\T_\NL$, and for various values of $\K$ and of $\M$, we diffuse $\M$ randomly generated signals $\K$ times using $\T_\NL$.
            From the diffused signals, we compute the eigenvectors of the sample covariance matrix, and check if the eigenvalues of $\T_\NL$ are located in the polytope defined by these eigenvectors.
            \figref{convergenceOfEigenvectorsKM} depicts the ratio of times it is the case, for each combination of $\K$ and $\M$.
            
            The figure demonstrates that the number of diffusions has some imporance in the process.
            Too small values of $\K$ encode too little information on the diffusion matrix in the signals, but values that are too high concentrate the eigenvalues too much around $0$.
            This corroborates that, as the eigenvalues concentrate around $0$, it is necessary to have a higher value of $\M$ to achieve the same precision.
            \stopNewText
            
        %
        
    %


    \section{Strategies for selecting a diffusion matrix}
    \label{strategies}
        
        As stated in \secref{admissibleMatrices}, inferring a valid diffusion matrix, in the sense that it can explain the relationships among signal entries through a diffusion process, reduces to selecting a point in the polytope.
        Since it contains an infinite number of possible solutions, one needs to introduce additional selection criteria in order to favor desired properties of the retrieved solution.
        
        \newText
        Note that the polytope describes a set of diffusion matrices as introduced in \defref{diffusionMatrix}.
        Given a diffusion matrix $\rec\T$ selected from the polytope, unless the degrees of the vertices are known, there is no possibility in the general case to retrieve the corresponding adjacency matrix.
        However, in the particular case when the associated adjacency matrix is binary, one can just threshold $\rec\T$ at $0$, setting its non-null entries to $1$.
        \stopNewText
        
        In this section, we first propose to illustrate the selection of points in the polytope, using two criteria: simplicity of the solution, and sparsity.
        In the first case, we aim at retrieving a diffusion matrix that has an empty diagonal.
        In the second case, we aim at recovering a sparse diffusion matrix.
        \newText
        Additionally, we introduce a third method that performs differently from the two other methods.
        Numerous graph inference techniques have been developed to obtain a graph from signals, with various priors.
        While most of them do not require the retrieved matrices to share the eigenvectors of the covariance matrix, it may still be interesting to evaluate whether these matrices are close enough to the polytope.
        If one can select a point in the polytope that is \emph{close} to the solution of a given method, while keeping the properties enforced by the associated priors, we obtain a new selection strategy.
        For this reason, we introduce in this section a method to adapt the solutions of other methods to stationary signals.
        \stopNewText
        
        \subsection{Selecting a diffusion matrix under a simplicity criterion}
        \label{simpleStrategy}
            
            The first criterion we consider to select a point in the polytope is simplicity of the solution.
            In other words, we want to encourage the retrieval of a set \rec\eigval{} of eigenvalues that, jointly with the eigenvectors \eigvec{} of the covariance matrix, produce a diffusion matrix that has an empty diagonal.
            Such a matrix represents a process that maximizes the diffusion of a signal evolving on it, and does not retain any of its energy.
            
            As shown in \secref{admissibleMatrices}, since we are considering a diffusion matrix as defined in \defref{diffusionMatrix}, the polytope of solutions is defined by inequality constraints \eqref{inequationsPositivity} that each enforce the positivity of an entry in the matrix to recover.
            A consequence is that if the matrix to be retrieved contains any null entry, then the point we want to select lies on an edge or a face of the polytope, since at least one inequality constraint holds with equality.
            Enforcing simplicity of the solution is therefore equivalent to selecting a point in the polytope that is located at the intersection of at least \N{} constraints.
            Using this observation and the fact that the trace of a matrix is equal to the sum of its eigenvalues, retrieving the eigenvalues that enforce simplicity of the corresponding matrix reduces to solving a linear programming problem, stated as follows:
            \begin{equation}
                \begin{array}{l}
                    \rec{\el_1}, \dots, \rec{\el_\N} = \displaystyle\arg\min_{\el_1, \dots, \el_\N}~~\displaystyle\sum_{i = 1}^\N \el_i \\
                    \st{} \left\{\begin{array}{l}
                        \eqref{inequationsPositivity} \\
                        \forall i \in \intInterval{1}{\N} : \el_i \in \interval{-1}{1} \\
                        \el_1 = 1
                    \end{array}\right. \;,
                \end{array}
                \label{problemSimple}
            \end{equation}
            where the two last constraints impose a scale factor.
            
            Equation \eqref{problemSimple} is a linear program for which it is known that polynomial-time algorithms exist.
            The main bottleneck of this method is the definition of the $\frac{\N(\N+1)}{2}$ linear constraints in \eqref{inequationsPositivity}, that are computed in $\mathcal{O}(\N^3)$ time and space.

        \subsection{Selecting a diffusion matrix under a sparsity criterion}
        \label{sparseStrategy}
            
            In many applications one may believe the graph underlying the observations is sparse.
            Similar to the case when trying to recover a simple graph, finding a sparse admissible solution can be formulated as finding a point at the intersection of multiple linear constraints.
            To find a sparse solution, we seek the set of admissible eigenvalues for which the maximum number of constraints in \eqref{inequationsPositivity} are null.
            This reduces to minimizing the \lNorm{0} norm of the solution, which is an NP-hard problem \cite{Amaldi1998}.
            
            A common approach to circumvent this problem is to approximate the minimizer of the \lNorm{0} norm by minimizing the \lNorm{1} norm instead \cite{Chen2001, Gribonval2003, Rinaldi2009}.
            In our case, we use the \LNorm{1}{1} matrix norm, which is the sum of all entries, since they are all positive.
            In this section, we adopt this approach and consider again a linear programming problem as follows:
            \begin{equation}
                \begin{array}{l}
                    \rec{\el_1}, \dots, \rec{\el_\N} = \displaystyle\arg\min_{\el_1, \dots, \el_\N}~\tr{\one_\N} \eigvec \begin{psmallmatrix} \el_1 & 0 & 0 \\ 0 & \dots & 0 \\ 0 & 0 & \el_\N \end{psmallmatrix} \tr{\eigvec} \one_\N \\
                    \st{} \left\{\begin{array}{l}
                        \eqref{inequationsPositivity} \\
                        \forall i \in \intInterval{1}{\N} : \el_i \in \interval{-1}{1} \\
                        \el_1 = 1
                    \end{array}\right. \;,
                \end{array}
                \label{problemSparse}
            \end{equation}
            where $\one_\N$ is the vector of \N{} entries all equal to one.

        \newText
        \subsection{Adaptation of other strategies to stationary signals}
        \label{regularizationStrategy}
            
            The two methods introduced before consist in selecting a point in the polytope, given simplicity or sparsity priors.
            In this section, we take a different point of view.
            Many graph inference techniques exist in the literature (see \secref{relatedWork}), all enforcing different properties of the graph that is retrieved.
            However, most of them do not impose the eigenvectors of the inferred solution to match those of the covariance matrix.
            The idea here is to adapt these solutions to stationary signals.
            
            To do so, let us consider an inference method $m$ providing an adjacency or a Laplacian matrix from a set of signals $\X = \orderedSet{\x_1, \dots, \x_\M}$.
            Let $\T_m$ be a diffusion matrix associated with the inferred matrix (for example, the one derived from the normalized Laplacian).
            Let $\eigvec_{\T_m}$ be the eigenvectors of $\T_m$, and let $\eigvec_\cov$ be the eigenvectors of the covariance matrix.
            
            The idea here is to consider $\T_m$ as if it were expressed in the eigenbasis of $\eigvec_\cov$, to check whether it belongs or not to the polytope of admissible matrices.
            In other words, we want to find a matrix $\A_m$ such that $\T_m = \eigvec_\cov \A_m \eigvec_\cov^\top$.
            Using the fact that $\eigvec_\cov$ forms an orthonormal basis, we have $\A_m = \eigvec_\cov^\top \T_m \eigvec_\cov$.
            Unless $\eigvec_{\T_m}$ and $\eigvec_\cov$ are the same, $\A_m$ is not necessarily a diagonal matrix.
            Therefore, $\A_m$ lies in a space of dimension $\N^2$, while the polytope is defined by \N{} variables.
            
            Let us call $\eigval_m = \orderedSet{\el_{m_1}, \dots, \el_{m_\N}}$ the vector of elements on the diagonal of $\A_m$.
            Since the polytope of admissible diffusion matrices is defined in $\setR^\N$, $\eigval_m$ is the point of this set that forms the best estimate for $\A_m$, defined in $\setR^{\N^2}$, after dimensionality reduction.
            In other words, $\eigval_m$ is the orthogonal projection of $\A_m$ in the polytope.
            If $\eigval_m$ does not belong to the polytope of admissible diffusion matrices characterized by $\eigvec_\cov$, then the method $m$ provides a solution that does not satisfy the conditions to be a diffusion process.
            To find the point in the polytope that is the closest to $\eigval_m$ in the sense of the Euclidean norm, we solve the following problem:
            \begin{equation}
                \begin{array}{l}
                    \rec{\eigval_m} = \arg\displaystyle\min_{\eigval \in \setR^\N} \| \eigval - \eigval_m \|_2 \\
                    ~~\st{} \left\{\begin{array}{l}
                        \eqref{inequationsPositivity} \\
                        \forall i \in \intInterval{1}{\N} : \veci{\eigval}{i} \in \interval{-1}{1} \\
                        \veci{\eigval}{1} = 1
                    \end{array}\right. \;.
                \end{array}
                \label{problemClosest}
            \end{equation}
            
            The solution to \eqref{problemClosest} gives us a set of eigenvalues $\rec{\eigval_m} = \orderedSet{\rec{\el_{m_1}}, \dots, \rec{\el_{m_\N}}}$ that represent the best approximation of $\A_m$ when restricting the search to admissible diffusion matrices.
            Therefore, the matrix $\rec{\T_m} = \eigvec_\cov \begin{psmallmatrix} \rec{\el_{m_1}} & 0 & 0 \\ 0 & \dots & 0 \\ 0 & 0 & \rec{\el_{m_\N}} \end{psmallmatrix} \eigvec_\cov^\top$ is the adaptation of the solution of method $m$ to stationary signals.
            We measure the distance between $\A_m$, the projection of $\T_m$ in the space defined by $\eigvec_\cov$, and $\rec{\eigval_m}$, the closest point in the polytope, as follows:
            \begin{equation}
                d(\A_m, \rec{\eigval_m}) \triangleq \left\| \A_m - \begin{pmatrix} \rec{\el_{m_1}} & 0 & 0 \\ 0 & \dots & 0 \\ 0 & 0 & \rec{\el_{m_\N}} \end{pmatrix} \right\|_F
                \;,
                \label{correctionDistance}
            \end{equation}
            where $\| \cdot \|_F$ is the Frobenius norm.
            
\begin{figure}
                \centering
                \includegraphics[width=0.85\linewidth]{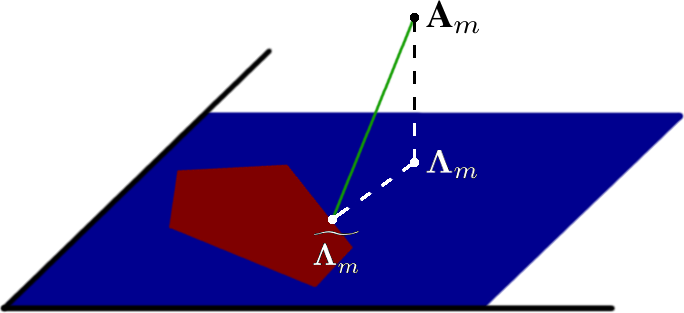}
                \caption
                {
                    Correction of the result of an inference method $m$ to match the stationarity hypothesis on the observed signals.
                    The eigenvalues of the result of $m$ are expressed in the space defined by $\eigvec_\cov$ as a matrix $\A_m$.
                    Then, $\A_m$ is approximated by $\eigval_m$, its orthogonal projection in the space of the polytope.
                    The closest point in the polytope (in the sense of the Euclidean norm), $\rec{\eigval_m}$, is then found by solving \eqref{problemClosest}.
                    Finally, the distance between $\A_m$ and its estimate in the polytope is given by the measure in \eqref{correctionDistance}.
                    This corresponds to the norm of the vector in green.
                }
                \label{regularizationFigure}
            \end{figure}

            \figref{regularizationFigure} summarizes provides a graphical illustration of the various steps to compute this distance.

        \stopNewText
        
    %


    \section{Numerical experiments}
    \label{experiments}
        
        To be able to evaluate reconstruction performance of the methods presented in \secref{strategies}, we first need to design experimental settings.
        This section introduces a generative model for graphs and signals, and evaluates the performance of the methods \emph{Simple} and \emph{Sparse}.
        \newText
        The regularization method introduced in \secref{regularizationStrategy} is evaluated as a means to select a matrix representing best some given signals, among a set of possible matrices.
        \stopNewText
        \secref{experimentsReal} presents additional experiments and comparisons with other methods from the literature on a non-synthetic dataset of temperatures in Brittany.
        
        \newText
        Our experiments show that the \emph{Simple} method succeeds in recovering the ground truth matrix from signals diffused on it, provided that the number of signals is high enough.
        The \emph{Sparse} method, while not being able to retrieve the ground truth matrix, infers a matrix that has a lower \LNorm{1}{1} norm than the matrix yielding the polytope.
        Additionally, we show that the regularization method allows the selection of the ground truth diffusion matrix from a set of candidate matrices, even for a small number of signals.
        Finally, comparison with other inference methods on a dataset of temperatures show that the methods \emph{Simple} and \emph{Sparse} return the best solutions with respect to their objectives, and that the regularization strategy applied to methods favorizing smoothness of signals yields a diffusion matrix on which signals are relatively smooth.
        \stopNewText
        
        \subsection{Generative model for graphs and signals}
        \label{generativeModel}

            In our experiments, we consider randomly generated graphs, produced by a random geometric model.
            These are frequently used to model connectivity in wireless networks \cite{Nekovee2007}.
            
            \begin{definition}[Random geometric graph]
                A random geometric graph of parameter \R{} is a graph built from a set of \N{} uniformly distributed random points on the surface of a unit 2-dimensional torus, by adding an edge between those being closer than \R{} according to the geodesic distance $\d(i, j)$ on the torus.
                We then add a weight on the existing edges that is inversely proportional to the distance separating the points.
                Here, we choose to use the inverse of $\d(i, j)$.
                The adjacency matrix \W{} of such a graph is defined by:
                \begin{equation}
                        \matij{\W}{i}{j} \triangleq \left\{
                                                        \begin{array}{cl}
                                                            \frac{1}{\d(i, j)} & \eqif \d(i, j) < \R \eqand i \neq j \\
                                                            0 & \eqotherwise
                                                        \end{array}
                                                    \right. \;.
                \end{equation}
                \label{randomGeometric}
            \end{definition}
            
            Note that, by construction, such graphs are simple and relatively sparse.
            Therefore, we expect methods using such selection criteria to be able to retrieve them.
            
            \newText
            In some of our experiments, we also consider random graphs generated by an Erdős-Rényi model \cite{Erdos1959}, in which two vertices are linked with a given probability independently from each other.
            Such graphs are defined as follows:
            
            \begin{definition}[Erdős-Rényi graph]
                An Erdős-Rényi graph of parameter \P{} is a graph where each edge exists with probability $P$ independently from each other.
                The adjacency matrix \W{} of such a graph is defined by:
                \begin{equation}
                        \matij{\W}{i}{j} \triangleq \left\{
                                                        \begin{array}{cl}
                                                            1 & \text{with probability~} \P \\
                                                            0 & \text{with probability~} 1 -\P
                                                        \end{array}
                                                    \right. \;.
                \end{equation}
                \label{erdosRenyi}
            \end{definition}
            \stopNewText
            
            Once a graph is generated using the model presented above, and given a number $\M$ of signals to produce, signals verifying our settings are created as follows:
            \begin{enumerate}
                \item Create \Y{} a $\N \times \M$ matrix with \iid{} entries.
                      We denote by \veci{\Y}{i} the $i^\text{th}$ column of \Y.
                      In these experiments, entries of \Y{} are drawn uniformly.
                \item Create \k{} a vector of \M{} \iid{} integer entries, comprised in the interval $\intInterval{1}{10}$.
                      These values are chosen not too high in reaction to a remark in \secref{gft}, not to obtain signals that are already stable.
                      In these experiments, entries of \k{} are also drawn uniformly.
                \item Compute the diffusion matrix $\T_\NL$ associated with \G.
                \item Create \X{} a $\N \times \M$ matrix of signal as follows: $\forall i \in \intInterval{1}{M} : \veci{\X}{i} \triangleq \T_\NL^{\veci{\k}{i}} \veci{\Y}{i}$.
            \end{enumerate}
            
            \newText
            Once these four steps are performed, the objective becomes: given \X{} and some criteria on the graph to retrieve, infer an estimate \rec\T{} for the diffusion matrix of the signals.
            To summarize the previous sections, we proceed as follows:
            \begin{enumerate}
                \item Find $\rec\eigvec$, an estimate for the eigenvectors of the diffusion matrix.
                      Here, this is done by computing the eigenvectors of the sample covariance matrix.
                \item Using $\rec\eigvec$, compute the constraints in \eqref{inequationsPositivity} that define the polytope of solutions.
                \item Select a point from the polytope, using one of the strategies in \secref{strategies}.
            \end{enumerate}
            \stopNewText

        \subsection{Error metrics}
        \label{errorMeasurements}

            To be able to evaluate the reconstruction error for our techniques, we use multiple metrics.
            Let \T{} be the ground truth diffusion matrix, with eigenvalues $\eigval = \orderedSet{\el_1, \dots, \el_\N}$, and let \rec\T{} be the one that is recovered using the assessed technique, with eigenvalues $\rec\eigval = \orderedSet{\rec{\el_1}, \dots, \rec{\el_\N}}$.
            
            The first metric we propose is the mean error per reconstructed entry (\MEPRE):
            \begin{equation}
                \MEPRE(\T, \rec\T) \triangleq \frac{1}{\N} \left\| \frac{\T}{\norm{\T}{F}} - \frac{\rec\T}{\norm{\rec\T}{F}}\right\|_F \;.
                \label{MEPRE}
            \end{equation}
            This quantity measures the mean error for all entries in the reconstructed matrix, where we first normalize \T{} and \rec\T{} using their Frobenius norm \norm{\cdot}{F} to avoid biases related to scale.
            
            The second metric we propose is the reconstruction error of the powered retrieved eigenvalues (\REPRE).
            We define it as the Euclidean distance between the $\K$th-power of the ground truth vector of eigenvalues $\eigval^\K$ and the recovered ones \rec\eigval{}, for the best value of \K{} possible:
            \begin{equation}
                \REPRE(\eigval, \rec\eigval) \triangleq \displaystyle \min_{\K \in \setR} \frac{1}{\N}\left\|\frac{\eigval^\K}{\norm{\eigval^\K}{\infty}} - \frac{\rec\eigval}{\norm{\rec\eigval}{\infty}}\right\|_2 \;.
                \label{REPRE}
            \end{equation}
            Here, the normalization using \norm{\cdot}{\infty} comes from the constraint in \secref{admissibleMatrices} that the highest eigenvalue should be equal to $1$.
            Therefore, it imposes a scale on the set of eigenvalues.
            Also, we divide the error by \N{} to make it independent of the number of vertices.
            
            Since in the limit case $\cov = \T^{2\K}$, for an unknown $\K \in \setR$, then the algorithms should be able to recover at least a power of \T.
            Indeed, there is absolutely no way to distinguish for example one step of diffusion of a signal \x{} using $\T^{2\K}$ (\ie, $\T^{2\K} \x$) and two diffusion steps of \x{} using $\T^\K$ (\ie, $\left(\T^\K\right)^2 \x$).
            A consequence is that, if the algorithm cannot fully retrieve \T, a power of \T{} should also be an acceptable answer.
            

            
            These two metrics provide information on the ability of methods to infer a diffusion matrix that is close to the ground truth one, or one of its powers.
            More classical metrics are also considered to evaluate whether the most significant entries of the inferred matrices correspond to existing edges in the ground truth graph.
            Since such metrics are defined for binary, relatively sparse graphs, we evaluate them on thresholded versions of the inferred matrix.
            Entries of the inferred matrix that are above this threshold $t$ are set to $1$, and others are set to $0$.
            The resulting matrix is denoted in the following equations as $\rec\T_{t}$.
            To find the optimal value, we perform an exhaustive search among all possible thesholds, and keep the one that maximizes the F-measure, defined below.
            
            The first metric we consider is the \emph{Precision}, measuring the fraction of relevant edges among those retrieved.
            \begin{equation}
                \precision(\T, \rec\T_{t}) \triangleq \resizebox{0.6\linewidth}{!}{$\frac{\#\left\{(i,j) | \rec\T_{t}(i, j) > 0 \text{~and~} \T(i, j) > 0 \right\}}{\|\rec\T_{t}\|_{0, 1}} \;,$}
                \label{precision}
            \end{equation}
            where $\|\cdot\|_{0, 1}$ denotes the \LNorm{0}{1} matrix norm, that counts the number of non-null entries.
            
            A second metric we consider is the \emph{Recall}, that measures the fraction of relevant edges effectively retrieved.
            \begin{equation}
                \recall(\T, \rec\T_{t}) \triangleq \resizebox{0.65\linewidth}{!}{$\frac{\#\left\{(i,j) | \rec\T_{t}(i, j) > 0 \text{~and~} \T(i, j) > 0 \right\}}{\|\T\|_{0, 1}} \;,$}
                \label{recall}
            \end{equation}
            
            Both metrics are often combined into a single one, called \emph{F-measure}, that can be considered as a harmonic mean of precision and recall.
            \begin{equation}
                    \fmeasure(\T, \rec\T_{t}) \triangleq 2 \resizebox{0.5\linewidth}{!}{$\frac{\precision(\T, \rec\T_{t}) \cdot \recall(\T, \rec\T_{t})}{\precision(\T, \rec\T_{t}) + \recall(\T, \rec\T_{t})} \;.$}
                \label{fMeasure}
            \end{equation}
            
            Note that in practical cases, the optimal threshold is not available, and depends on a desired sparsity of the inferred matrix.
            In the following experiments, we show the compromise between true positive edges and false positive edges for all possible thresholds using ROC curves.

        \newText
        \subsection{Performance of the Simple method}
        \label{performanceSimple}
        \stopNewText
            
            In the situation when the eigenvectors are not available, due to a limited number of signals, recovery methods must use estimated eigenvectors.
            Linear programming problems introduced in \secref{strategies} must then be solved on a polytope defined by noisy eigenvectors.
            We have previously shown in \secref{estimatedEigenvectors} that increasing the number of signals allows this approximate polytope to be more precise.
            The following experiments evaluate the quality of the solutions retrieved by the two methods introduced in \secref{simpleStrategy} and \secref{sparseStrategy}.

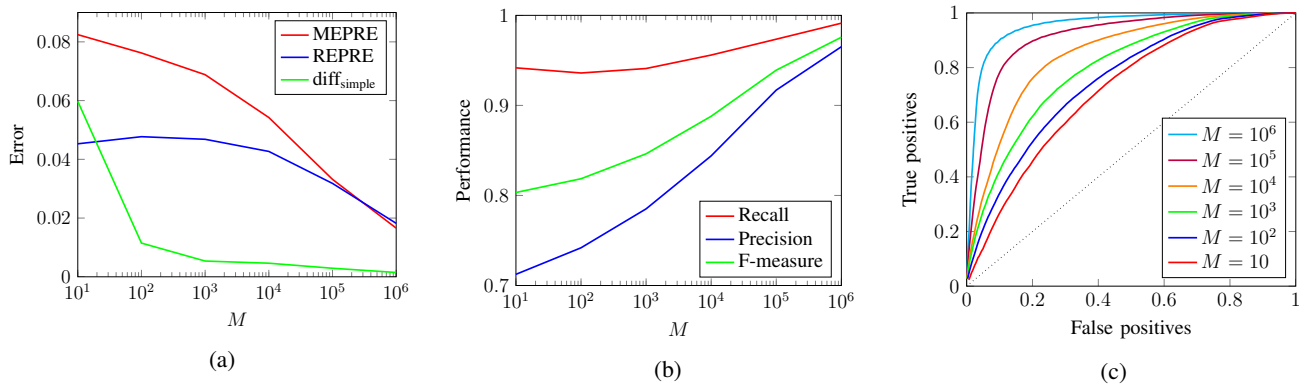
\begin{figure*}
                \centering
                \begin{subfigure}{0.32\textwidth}
                    \resizebox{0.96\linewidth}{!}{\large{\input{resultsErrorsSimple.tex}}}
                    \caption{}
                \end{subfigure}
                \begin{subfigure}{0.32\textwidth}
                    \resizebox{0.96\linewidth}{!}{\large{\input{resultsErrorsSimple2.tex}}}
                    \caption{}
                \end{subfigure}
                \begin{subfigure}{0.32\textwidth}
                    \resizebox{0.96\linewidth}{!}{\large{\input{resultsROCSimple.tex}}}
                    \caption{}
                \end{subfigure}
                \caption
                {
                    Image (a) depicts the mean \MEPRE{} and \REPRE{} measurements of the solutions retrieved by the \emph{Simple} method, for $\M \in \set{10^i}_{1 \leq i \leq 6}$ signals.
                    Additionally, $\diff_\text{simple}$ shows the distance to the ground truth solutions in terms of trace value, which is the objective function of problem presented in \eqref{problemSimple}.
                    Image (b) shows the results in terms of edge reconstruction by studying the recall, precision and F-measure using the binarized thresholded matrix $\rec\T_{t}$.
                    Finally, image (c) depicts the ROC curves, that show the compromise between true positive edges and false positive edges when varying the threshold.
                    All tests were performed for $1000$ occurrences of random geometric graphs with parameters $\N = 10$ and $\R = 0.6$.
                }
                \label{resultsErrorsSimple}
            \end{figure*}

            \figref{resultsErrorsSimple} illustrates the convergence of the \emph{Simple} method to a solution, when the number of signals increases.
            In this experiment, we generate $1000$ random geometric graphs ($\N = 10$, $\R = 0.6$) and, for each of them, we diffuse $\M \in \set{10^i}_{1 \leq i \leq 6}$ signals using $\T_\NL$, as described in \secref{generativeModel}.
            For each configuration, we retrieve a diffusion matrix by solving the problem in \eqref{problemSimple} using the CVX \cite{Grant2014} package for MATLAB \cite{MATLAB2012}, with default parameters.
            Then, we compute the mean errors, and measure the distance to the ground truth solutions in terms of trace value, which is the objective function in \eqref{problemSimple}:
            \begin{equation}
                \diff{}_\text{simple}(\T_\NL, \rec\T) \triangleq \frac{1}{\N} \left(\trace(\rec\T) - \trace(\T_\NL)\right) \;,
                \label{differenceSimple}
            \end{equation}
            which in our case simplifies to $\diff_\text{simple}(\T_\NL, \rec\T) \triangleq \frac{1}{\N} \trace(\rec\T)$ since by construction $\trace(\T_\NL) = 0$.
            
            These results show that both error measurements decrease --- except for very low values of \N{} due to the high variance --- as \M{} increases.
            As the approximate polytope converges to the ground truth one, the solution of \eqref{problemSimple} converges to the ground truth matrix $\T_\NL$ used to produce the signals.
            This is confirmed by performance metrics as well as the ROC curves, which indicate that the ground truth edges are recovered more successfully as the number of signals increases.
            
            As stated in \secref{generativeModel}, matrices generated using the random geometric model are by construction simple, and have therefore null traces.
            When defining the polytope using inequalities in \eqref{inequationsPositivity}, we enforce the positivity of all entries of the admissible matrices.
            Therefore, $\T_\NL$ is a matrix for which eigenvalues lie on a plane where $\el_1 + \el_2 + \el_3 = 0$.
            While the optimal solution may not be unique (see below), counter-examples to this uniqueness must respect particular constraints and are very unlikely to happen for random matrices as well as for non-synthetic cases.
            As a consequence, minimizing the sum of the eigenvalues as an objective function enforces the retrieval of the correct result in nearly all cases.
            This implies that the error measurements will certainly converge to $0$ as \M{} grows to infinity.
            This can easily be verified by replacing \rec\eigvec{} by \eigvec{} --- the eigenvectors of $\T_\NL$ --- in \eqref{inequationsPositivity} for the resolution of \eqref{problemSimple}.
            
            As stated above, the solution is not necessarily unique.
            Multiple matrices with a minimum trace can be found when there exists a frontier of the polytope along which all points have a sum that is minimal among all admissible vectors of eigenvalues.
            As an example, let us consider the $8 \times 8$ Hadamard matrix as a matrix of eigenvectors $\eigvec$:
            When defining the polytope associated with these eigenvectors using constraints in \eqref{inequationsPositivity}, we obtain for $\T(2,6)$ the following constraint:
            \begin{equation}
                \el_2 + \el_3 + C \geq 0 \;,
                \label{counterExampleEquation}
            \end{equation}
            where $C$ is some value that does not depend on $\el_2$ and $\el_3$.
            As a consequence, any point located on this particular plane corresponds to a matrix with identical sum of eigenvalues, leading to the same trace.
            When considering random matrices or non-synthetic cases, the case when there exists such a plane that is aligned with the objective is very unlikely.

        \newText
        \subsection{Performance of the Sparse method}
        \label{performanceSparse}
        \stopNewText
        
            In this section, we perform the same experiment as for the \emph{Simple} method, but solving the problem in \eqref{problemSparse} instead of \eqref{problemSimple}.
            Here, since we minimize the \LNorm{1}{1} norm as an objective function, we measure the mean difference between the sparsity of the retrieved matrices and the sparsity of the ground truth matrices $\T_\NL$, computed as follows:
            \begin{equation}
                \diff{}_\text{sparse}(\T_\NL, \rec\T) \triangleq \frac{1}{\N^2} \left(\|\rec\T\|_{1, 1} - \|\T_\NL\|_{1, 1}\right) \;.
                \label{differenceSimple}
            \end{equation}
            
            \newText
            For space considerations, the results are not detailed here.
            Contrary to the method for selecting a simple graph in \secref{simpleStrategy}, we have observed that the error and performance measurements stay approximately constant for all values of \M{} ($\MEPRE \approx 0.1$, $\REPRE \approx 6 \times 10^{-2}$, $\diff{}_\text{sparse} \approx -2 \times 10^{-2}$, $\fmeasure \approx 0.76$).
            Similar results were obtained for additional experiments conducted on random geometric graphs with different values of \R{} to assess different levels of sparsity.
            
            The results suggest that the method fails at recovering the matrix that was used for diffusing the signals, even when the number of observations is high.
            However, the negative difference indicates that the method does not fail at recovering a sparse graph, in the sense of the \LNorm{1}{1} norm.
            Graphs generated using the random geometric model, while being sparse by construction, are not necessarily the sparsest within the associated admissible set, especially when considering the \LNorm{1}{1} norm.
            This implies that, as the number of signals increases, the method in fact converges to the sparsest solution in the polytope, although it is in most cases not the matrix we started from.
            Replacing \rec\eigvec{} by \eigvec{} --- the eigenvectors of $\T_\NL$ --- in \eqref{inequationsPositivity} for the resolution of \eqref{problemSparse} confirms that there exist sparser solutions than the ground truth graph.

            Note that in their work, Segarra \etal{} \cite{Segarra2016b} also use the \LNorm{1}{1} norm minimization as an objective, and suceed in retrieving the ground truth graph.
            This is the case because they have additional constraints that enforce the solution to be simple.
            Therefore, the set of solutions is a lot smaller, and the solution is most likely unique.
            \stopNewText

        \newText
        \subsection{Impact of the parameters}
        \label{performanceSimpleSparseParameters}
        
            The quality of the solutions inferred by the methods assessed above mostly depends on how close the eigenvectors of the sample covariance matrix are from the ground truth ones.
            Noticing that the density of eigenvalues in the interval $\interval{-1}{1}$ increases with \N{} for any diffusion matrix, it follows that larger graphs have eigenvalues pairwise closer than for smaller graphs.
            In this respect, \secref{estimatedEigenvectors} tells us that the number of signals necessary for a precise estimate of the covariance matrix needs to be higher for larger graphs.
            This raises the question of the scalability of this method.
            
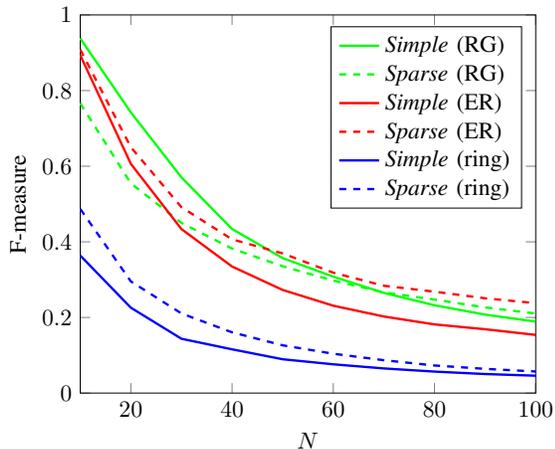
\begin{figure}
                \centering
                \resizebox{0.85\linewidth}{!}{\input{fmeasureVaryingN.tex}}
                \caption
                {
                    F-measure scores obtained for the \emph{Simple} and \emph{Sparse} methods, for various families of graphs and for $\M = 10^5$ signals, as a function of the graph order.
                    Tests were performed on $1000$ occurrences of each family of graph, for each technique.
                }
                \label{fmeasureVaryingN}
            \end{figure}

            For a fixed value of $\M = 10^5$ signals, we study the performance of the \emph{Simple} and \emph{Sparse} methods on random geometric graphs of orders ranging from $\N = 10$ to $\N = 100$ vertices.
            For each value of $\N$, we set $\R \varpropto \frac{1}{\sqrt{N}}$ so that the value of $\N$ does not impact the average neighborhood of each vertex.
            This value is chosen in accordance with the experiments performed earlier in this section.
            \figref{fmeasureVaryingN} depicts the F-measure performance measurements obtained for the \emph{Simple} and \emph{Sparse} methods, for $1000$ occurrences of random geometric graphs with the previously described settings.
            Additionally, we plot in this figure the results obtained for other families of graphs, namely Erdős-Rényi graphs with $\P \varpropto \frac{\log \N}{\N}$, and the ring graph of \N{} vertices.
            
            These measurements confirm that the inference methods need a large quantity of signals to work properly.
            Additionally, it appears that the family of graphs has an impact on the reconstruction performance.
            The ring graph has repeated eigenvalues, which has a strong impact on the convergence of the eigenvectors of the sample covariance matrix to those of the real covariance matrix (see \secref{estimatedEigenvectors} and \cite{Anderson1963}).
            However, this is a marginal case, since real-weighted graphs almost surely have distinct eigenvalues, as illustrated by the studies on random geometric and Erdős-Rényi graphs.
            Another parameter that has importance on these experiments is the number of diffusions of signals before observations, represented by the vector \k.
            This has been illustrated in \secref{estimatedEigenvectors}.

        \stopNewText
        
        \newText
        \subsection{Application of regularization to graph hypothesis testing}
        \label{regularizationStrategyEvaluation}
            
            To evaluate the practical interest of the regularization strategy introduced in \secref{regularizationStrategy}, let us consider the situation where some signals are observed, and various diffusion matrices are provided by inference methods to explain these signals.
            The objective is to determine which of the proposed solutions matches the signals best under a stationarity assumption.
            
            In the following experiment, we proceed as follows: let $\orderedSet{\T_1 \dots \T_{20}}$ be a set of $20$ diffusion matrices corresponding to graphs of $\N = 10$ vertices, equally divided into random geometric graphs (with \R{} drawn uniformly in $\interval{0.2}{0.6}$) and Erdős-Rényi graphs (with \P{} drawn uniformly in $\interval{0.2}{0.6}$).
            For each of these matrices, let us diffuse $\M$ random signals as detailed in \secref{experiments} to obtain observations $\orderedSet{\X_1 \dots \X_{20}}$, where $\X_i$ is the set of signals obtained after diffusion by $\T_i$.
            From these sets, we can compute the eigenvectors of the sample covariance matrices $\orderedSet{\rec{\eigvec_1} \dots \rec{\eigvec_{20}}}$.
            
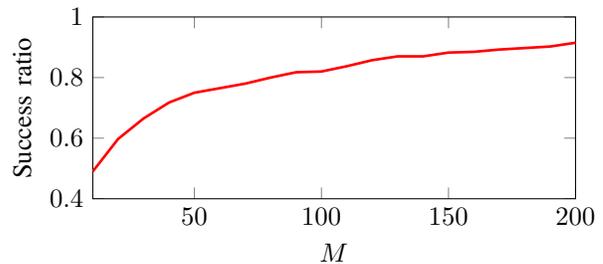
\begin{figure}
                \centering
                \input{graphSignalsAdequation.tex}
                \caption
                {
                    Ratio of times when the diffusion matrix $\T_i$ was chosen by the algorithm as the most adapted to signals $\X_i$ among a set of $20$ possible diffusion matrices, for $i \in \intInterval{1}{20}$.
                    Mean results for $100$ iterations of the experiment.
                }
                \label{graphSignalsAdequation}
            \end{figure}

            For each matrix of eigenvectors $\rec{\eigvec_i}$, $i \in \intInterval{1}{20}$, and for each diffusion matrix $\T_j$, $j \in \intInterval{1}{20}$, we compute the distance between the polytope yielded by $\rec{\eigvec_i}$ and the projection of $\T_j$ in the space of the polytope (see \secref{regularizationStrategy}) using \eqref{correctionDistance}.
            The graph that minimizes the distance is then selected as the most appropriate.
            \figref{graphSignalsAdequation} depicts the ratio of times $\T_i$ is selected as the most appropriate diffusion matrix when considering signals $\X_i$, for various values of \M.
            
            The results show that the regularization strategy selects the matrix used to diffuse the signals in most cases, even when \M{} is low.
            Additional experiments were performed for larger graphs, and similar results were observed.
            Also, increasing the number of signals eventually leads to a selection of the correct diffusion matrix in all cases.
            This experiment illustrates that the regularization strategy introduced in \secref{regularizationStrategy} can be successfully used to select the graph that is the most adapted to given signals among a set of candidates.
            
            An interesting direction for future work includes evaluation of the performance of the method when considering selection of the most adapted matrix from noisy versions of $\T_i$.

        \stopNewText
        
    %
    
    
    \section{Evaluation of inference methods on a dataset}
    \label{experimentsReal}
        
        The two methods introduced in \secref{simpleStrategy} and \secref{sparseStrategy} present solutions to infer a graph from signals, while ensuring that it is compliant with our diffusion prior.
        While other methods from the literature do not clearly impose this prior, evaluating whether they provide solutions that match a diffusion assumption is interesting, as it would provide additional selection strategies for admissible diffusion matrices if it is the case.
        \newText
        This section explores the application of the regularization strategy in \secref{regularizationStrategy} to the method of Kalofolias \cite{Kalofolias2016}, and shows that it provides matrices that do not belong to the polytope of solutions.
        The closest point in the polytope is considered, and evaluation on a dataset shows that the result \newText has interesting similarities with the original matrix.
        
        Throughout this section, we study an open dataset\footnote{In http://data.gouv.fr.} of temperature observations from 37 weather stations located in Brittany, France \cite{Girault2015a}.
        Our inference methods, as well as other existing methods, are evaluated on this dataset in terms of sparsity, trace of the solution, and smoothness.
        \stopNewText
        
        \subsection{Detailed evaluation of the method from Kalofolias}
        \label{kalofolias}
            
            The method from Kalofolias has two major qualities: it recovers a graph in a very short amount of time, and encourages smoothness of the solution, which can be a desirable property.
            \newText
            To evaluate whether the retrieved solution happens to match a diffusion process, let us consider the following experiment:
            \begin{enumerate}
                \item Let \G{} be a random geometric graph of $\N = 10$ vertices ($\R = 0.6$), and let $\T_\NL$ the diffusion matrix associated with its normalized Laplacian.
                      Using this matrix, we diffuse $\M = 10^6$ \iid{} signals as presented in \secref{generativeModel} to obtain a matrix \X.
                      Using Principal Component Analysis on \X{} \cite{Pearson1901}, we obtain $\rec\eigvec$, an estimate for the eigenvectors of $\T_\NL$.
                      This set of eigenvectors yields a polytope of admissible solutions.
                \item Then, we use the method from Kalofolias to infer a graph $\G_K$ from \X, and compute the associated matrix $\T_{\NL_K}$.
                      Since the log method from Kalofolias depends on parameters $\alpha$ and $\beta$, we keep the minimal distance obtained for values of $\alpha$ and $\beta$ ranging from $0.01$ to $2$, with a step of $10^{-2}$.
                      Equation \eqref{correctionDistance} gives us the distance between the polytope and the inferred solution.
                \item Additionally, we generate a random geometric graph $\G_R$ (independent from the ground truth one) from \X, using the same settings as for \G{} ($\N = 10$, $\R = 0.6$).
                      This gives us a baseline of how close a random graph with the same edges distribution can be to the ground truth one, and gives information on whether the results of Kalofolias are closer to the ground truth than a random matrix.
                      Again, \eqref{correctionDistance} measures the distance between the polytope and the associated matrix $\T_{\NL_R}$.
            \end{enumerate}
            
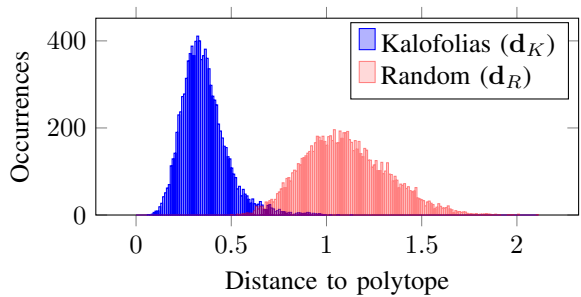
\begin{figure}
                \centering
                \input{resultsBothHistograms.tex}
                \caption
                {
                    Number of times a distance to the ground truth polytope was observed using either the method from Kalofolias \cite{Kalofolias2016} (${\bf d}_K$), or a random geometric graph (${\bf d}_R$).
                    Distances are grouped in bins of size $10^{-2}$.
                    Tests were performed for $10^5$ occurrences of graphs per method, with $\M = 10^6$ signals.
                }
                \label{resultsBothHistograms}
            \end{figure}

            We perform these three steps for $10^5$ occurrences of random geometric graphs.
            Let ${\bf d}_K$ be the vector of distances to the polytope obtained for each ground truth graph using the method of Kalofolias, and ${\bf d}_R$ the vector of distances to the polytope for the baseline random graphs.
            In \figref{resultsBothHistograms}, we plot a histogram of the number of times each distance was observed.
            
            \newText
            \stopNewText

            From these results, a first observation is that neither the methods from Kalofolias nor the random method ever returned a graph that was located in the polytope of solutions.
            Two direct interpretations of this result can be made: first, it implies that the set of admissible matrices per ground truth graph is small relatively to the set of random graphs.
            Second, it implies that the method from Kalofolias does not succeed in recovering a graph that matches diffusion priors on the signals.
            
            \newText
            Mann-Whitney $U$ test \cite{Mann1947} on ${\bf d}_K$ and ${\bf d}_R$ shows that the distributions differ significantly ($U = 9.9813 \times 10^7$, $P < 10^{-5}$ two-tailed).
            \stopNewText
            This implies that the results obtained with the method from Kalofolias are most of the time closer to an admissible matrix than random solutions.
            This observation can be explained by the remarks in \secref{smoothness}.
            Diffusion of signals on a graph tends to smoothen them, as the low frequencies are attenuated slower than higher ones.
            Since the method of Kalofolias retrieves a graph on which signals are smooth, the observation that it provides solutions that are closer to the polytope than random solutions is quite natural.

\begin{figure}
                \centering
                \begin{subfigure}{0.48\columnwidth}
                    \centering
                    \includegraphics[width=0.92\linewidth]{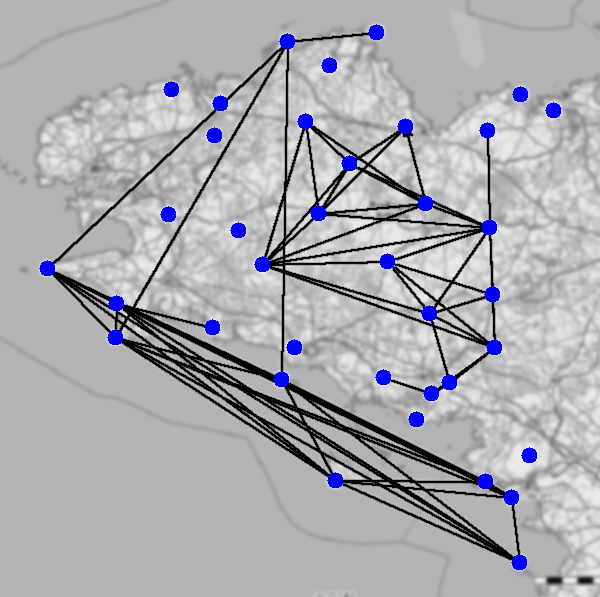}
                    \caption{}
                \end{subfigure}
                \begin{subfigure}{0.48\columnwidth}
                    \centering
                    \includegraphics[width=0.92\linewidth]{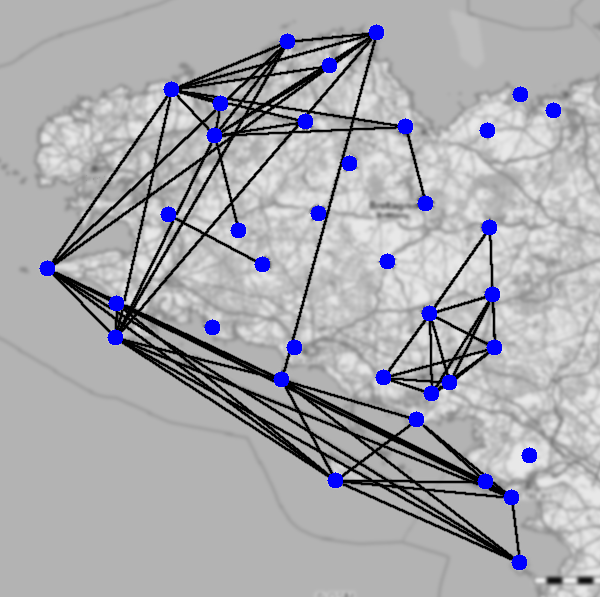}
                    \caption{}
                \end{subfigure}
                \caption
                {
                    Most significant connections in the adjacency matrix of the graph $\G_K$ retrieved by the method from Kalofolias (a), and most significant connections from the matrix associated with the closest point in the polytope, $\rec\T$ (b).
                }
                \label{bretagneKalofolias}
            \end{figure}

            The question is then whether the closest point to the retrieved solution in the polytope has interesting properties.
            Let us evaluate this solution on the dataset of temperatures.
            \figref{bretagneKalofolias} depicts the $10\%$ most significant connections in the adjacency matrix of the graph $\G_K$ retrieved by Kalofolias, as well as those of the matrix associated with the closest point in the polytope, $\rec\T = \rec\eigvec \rec\eigval \rec\eigvec^\top$, where $\rec\eigval$ is the solution of \eqref{problemClosest}.
            
            \newText
            The method from Kalofolias retrieves a matrix that has stronger connections between stations with similar locations.
            There is a strong connectivity among stations located on the south coast of Brittany.
            Stations located more in the land also tend to be linked to close inland stations.
            The regularized matrix appears to keep these properties: the strong links on the coasts still appear, and the result also still gives importance on the coastal versus inland aspect of the stations.
            Still, differences can be seen, as the regularized matrix appears to give more importance to the relations between stations on the north coasts.
            Such relations also exist in the original matrix, but are not depicted due to the threshold.
            
            When computing the total smoothness of the signals with both matrices, we obtain that the solution from Kalofolias has a higher value of smoothness (see \figref{comparisons}) than the closest point in the polytope.
            This implies that the signals are smoother on the approximate matrix than on the one recovered by the method of Kalofolias.
            This may seem counter-intuitive, since the solutions of the method by Kalofolias are not restricted to the polytope.
            However, the method from Kalofolias imposes inference of a matrix with an empty diagonal, which is not the case of the approximate one.
            These measurements, in addition to those below, suggest that inferring a graph using the method from Kalofolias, and considering the closest point in the polytope, is an interesting method to infer a valid graph on which signals are smooth.
            \stopNewText

        \subsection{Evaluation of graph inference methods on the dataset}
        \label{evalDataset}
            
            We have proposed in \secref{regularizationStrategy} a technique to find a valid matrix in the polytope that approximates the solution of any method.
            Therefore, it is possible to evaluate all methods in terms of properties, such as \LNorm{1}{1} sparsity, trace, or smoothness.
            \newText
            Since all methods do not impose the same scale on the inferred matrices, these quantities are computed for the inferred diffusion matrices after normalization such that their first eigenvalue equals one, as in the constraints in \defref{diffusionMatrix}.
            \stopNewText
            
            When applying our methods \emph{Simple} (\secref{simpleStrategy}) and \emph{Sparse} (\secref{sparseStrategy}), as well as those of Kalofolias \cite{Kalofolias2016}, Segarra \etal{} \cite{Segarra2016b} and the graphical lasso \cite{Friedman2008} on the dataset of temperatures, we have obtained the results in \figref{comparisons}.
            
            \newText
            \begin{figure}
                \centering
                \resizebox{\linewidth}{!}{
                    \begin{tabular}{|c||c|c|c|c|}
                        \hline
                        & polytope & $\LNorm{1}{1}$ & $\trace$ & $S(\X)$ \\
                        \hhline{|=||=|=|=|=|}
                        Simple & $\checkmark$ & $36.9974$ & ${\bf 0.0013}$ & $0.0551$ \\
                        \hline
                        Sparse & $\checkmark$ & ${\bf 36.9971}$ & $0.9093$ & $0.0585$ \\
                        \hline
                        Kalofolias \cite{Kalofolias2016} & $0.0313$ & $36.9979$ & $0$ & $0.0751$ \\
                        \hline
                        Kalofolias closest & $\checkmark$ & $36.9974$ & $0.0298$ & $0.0548$ \\
                        \hline
                        Segarra \etal{} \cite{Segarra2016b} & $0.0062$ & $36.9993$ & $1.97 \times 10^{-5}$ & $0.0245$ \\
                        \hline
                        Segarra \etal{} closest & $\checkmark$ & $36.9974$ & $0.0046$ & $0.0551$ \\
                        \hline
                        Graphical lasso \cite{Friedman2008} & $1.3730$ & $35.3539$ & $13.4977$ & $32.8421$ \\
                        \hline
                        Graphical lasso closest & $\checkmark$ & $36.9984$ & $13.3584$ & ${\bf 0.0335}$ \\
                        \hline
                    \end{tabular}
                }
                \caption
                {
                    Sparsity, trace and smoothness obtained for the dataset of temperatures.
                    Elements in bold denote the method performing best among those that return a solution located in the polytope.
                    If a method provides a solution that does not belong to the polytope, the distance to the closest point is indicated in the first column.
                    The last column indicates the total smoothness for all signals, \ie, $S(\X) \triangleq \sum_{\x \in \X} S(\x)$.
                }
                \label{comparisons}
            \end{figure}
            \stopNewText
            
            \newText
            First, we notice that the method from Segarra \etal{} \cite{Segarra2016b} returns a matrix that is at a distance of $0.0062$ from the polytope.
            As the polytope description is the same for their method and for ours, we would expect this distance to be $0$.
            This small difference comes from their implementation.
            In order to keep their equality constraints enforcing the elements in the polytope to have an empty diagonal, while coping with the noise in the eigenvectors, they allow small deviations from the polytope.
            They do not return a matrix ${\bf S}^*$ that shares the eigenvectors of the covariance matrix, but a matrix ${\bf \hat{S}}^*$ such that $\| {\bf S}^* - {\bf \hat{S}}^* \|_F \leq \varepsilon$.
            Here, experiments were performed for $\varepsilon = 10^{-3}$.
            For this reason, the matrix they return is located slightly outside of the polytope of solutions.
            When considering the closest point to this result in the polytope, it appears to be very close to the solution returned by the \emph{Simple} method.
            
            As expected, the \emph{Sparse} method recovers the matrix with the lowest \LNorm{1}{1} norm.
            It is also interesting to remark that the projection of the solution of the graphical lasso on the polytope is smoother than the projection of the solution obtained by the method from Kalofolias.
            This echoes the remark in \secref{graphicalLassoSection} that minimization of the quantity $\trace(\rec\cov \Th)$ in \eqref{graphicalLasso} tends to promote smoothness of the signals on the graph when $\Th$ is a Laplacian matrix, which appears to be encouraged by the regularization algorithm.
            Note that the method from Kalofolias infers a graph which projection on the polytope gets the second best smoothness score, while having a small trace.
            On the other hand, the solution infered by the graphical lasso appears to have most of its energy on the diagonal entries.
            This is confirmed by the traces of the matrices, both for the original solution and its approximate in the polytope.
            Therefore, these two solutions provide interesting ways to find a graph on which stationary signals are smooth, with different simplicity assumptions.
            \stopNewText
            
        %

    %


    \section{Conclusions}
    \label{conclusions}
        
        In this article, we have proposed a method for characterizing the set of matrices that may be used to explain the relationships among signal entries assuming a diffusion process.
        We have shown that they are part of a convex polytope, and have illustrated how one could choose a point in this particular set, given additional selection criteria such as sparsity of simplicity of the graph to infer.
        Finally, we have shown that most of other existing methods do not infer matrices that belong to the polytope of admissible solutions for stationarity signals, and have introduced a method to consider the closest valid matrix.
        An experiment was performed to illustrate that this particular method can be useful for graph hypothesis testing.
        
        Future directions based on this work are numerous.
        \newText
        First of all, reviewing the covariance estimation techniques is an interesting direction, as obtention of the eigenvectors of the covariance matrix is a cornerstone of our approach, and some techniques may provide such information more precisely than the sample covariance.
        \stopNewText
        We could also explore new strategies to select a point in the polytope, for example by enforcing the reconstruction of a binary matrix.
        Another interesting direction would then be to propose selection strategies that do not imply the full definition of the $\frac{\N(\N+1)}{2}$ constraints defining the polytope.
        \newText
        Finally, our immediate next work will be to complement our experiments on graph hypothesis testing, considering noisy versions of the candidate diffusion matrices.
        \stopNewText

%% file: inPolytopeVaryingMK.tex
\begin{tikzpicture}

    \begin{axis}[legend cell align={left}, width=6.5cm, height=4.9cm, xmin=1, xmax=20, ymin=0.3, ymax=1.0, xlabel=$K$, ylabel=Inclusion ratio, legend pos=outer north east]

        \addplot plot[mark=none, line width=1pt, color=orange]
        coordinates
        {
            (1, 0.9426)
            (2, 0.9670)
            (3, 0.9753)
            (4, 0.9811)
            (5, 0.9740)
            (6, 0.9332)
            (7, 0.8677)
            (8, 0.7703)
            (9, 0.6934)
            (10, 0.6278)
            (11, 0.6119)
            (12, 0.6178)
            (13, 0.6155)
            (14, 0.6090)
            (15, 0.6079)
            (16, 0.6230)
            (17, 0.6243)
            (18, 0.6140)
            (19, 0.6200)
            (20, 0.6144)
        };

        \addplot plot[mark=none, line width=1pt, color=purple]
        coordinates
        {
            (1, 0.8622)
            (2, 0.9208)
            (3, 0.9380)
            (4, 0.9553)
            (5, 0.9563)
            (6, 0.9299)
            (7, 0.8599)
            (8, 0.7703)
            (9, 0.6823)
            (10, 0.6030)
            (11, 0.5268)
            (12, 0.4809)
            (13, 0.4623)
            (14, 0.4467)
            (15, 0.4456)
            (16, 0.4494)
            (17, 0.4532)
            (18, 0.4475)
            (19, 0.4535)
            (20, 0.4495)
        };
        
        \addplot plot[mark=none, line width=1pt, color=green]
        coordinates
        {
            (1, 0.7240)
            (2, 0.8252)
            (3, 0.8640)
            (4, 0.8900)
            (5, 0.9107)
            (6, 0.8994)
            (7, 0.8446)
            (8, 0.7676)
            (9, 0.6659)
            (10, 0.5531)
            (11, 0.4553)
            (12, 0.3713)
            (13, 0.3485)
            (14, 0.3337)
            (15, 0.3341)
            (16, 0.3237)
            (17, 0.3310)
            (18, 0.3284)
            (19, 0.3338)
            (20, 0.3255)
        };
        
        \addplot plot[mark=none, line width=1pt, color=blue]
        coordinates
        {
            (1, 0.5673)
            (2, 0.6708)
            (3, 0.7344)
            (4, 0.7698)
            (5, 0.8059)
            (6, 0.8087)
            (7, 0.7867)
            (8, 0.7325)
            (9, 0.6560)
            (10, 0.5542)
            (11, 0.4416)
            (12, 0.3719)
            (13, 0.3427)
            (14, 0.3325)
            (15, 0.3238)
            (16, 0.3158)
            (17, 0.3243)
            (18, 0.3239)
            (19, 0.3265)
            (20, 0.3189)
        };

        \addplot plot[mark=none, line width=1pt, color=red]
        coordinates
        {
            (1, 0.3241)
            (2, 0.5368)
            (3, 0.5743)
            (4, 0.6086)
            (5, 0.6276)
            (6, 0.6516)
            (7, 0.6526)
            (8, 0.6425)
            (9, 0.5961)
            (10, 0.5093)
            (11, 0.4274)
            (12, 0.3640)
            (13, 0.3274)
            (14, 0.3249)
            (15, 0.3156)
            (16, 0.3163)
            (17, 0.3279)
            (18, 0.3192)
            (19, 0.3217)
            (20, 0.3171)
        };

        \legend{\small{$\M = 10^5$}, \small{$\M = 10^4$}, \small{$\M = 10^3$}, \small{$\M = 10^2$}, \small{$\M = 10$}}

    \end{axis}

\end{tikzpicture}

%% file: resultsErrorsSimple.tex
\begin{tikzpicture}

    \begin{semilogxaxis}[legend cell align={left}, xmin=10, xmax=1000000, ymin=0.0, ymax=0.09, scaled ticks=false, tick label style={/pgf/number format/fixed}, xlabel=$M$, ylabel=Error, legend pos=north east]

        \addplot plot[mark=none, line width=1pt, color=red]
        coordinates
        {
            (10, 0.08246)
            (100, 0.07622)
            (1000, 0.06882)
            (10000, 0.05424)
            (100000, 0.03318)
            (1000000, 0.01657)
        };

        \addplot plot[mark=none, line width=1pt, color=blue]
        coordinates
        {
            (10, 0.045287)
            (100, 0.047713)
            (1000, 0.046824)
            (10000, 0.042689)
            (100000, 0.031797)
            (1000000, 0.018206)
        };

        \addplot plot[mark=none, line width=1pt, color=green]
        coordinates
        {
            (10, 0.059752)
            (100, 0.011479)
            (1000, 0.005365)
            (10000, 0.004631)
            (100000, 0.002913)
            (1000000, 0.001475)
        };

        \legend{MEPRE, REPRE, diff$_\text{simple}$}

    \end{semilogxaxis}

\end{tikzpicture}

%% file: resultsErrorsSimple2.tex
\begin{tikzpicture}

    \begin{semilogxaxis}[legend cell align={left}, xmin=10, xmax=1000000, ymin=0.7, ymax=1.0, xlabel=$M$, ylabel=Performance, legend pos=south east]

        \addplot plot[mark=none, line width=1pt, color=red]
        coordinates
        {
            (10, 0.941706)
            (100, 0.935990)
            (1000, 0.940972)
            (10000, 0.955922)
            (100000, 0.973545)
            (1000000, 0.991314)
        };

        \addplot plot[mark=none, line width=1pt, color=blue]
        coordinates
        {
            (10, 0.712407)
            (100, 0.741873)
            (1000, 0.785115)
            (10000, 0.843992)
            (100000, 0.917016)
            (1000000, 0.965159)
        };

        \addplot plot[mark=none, line width=1pt, color=green]
        coordinates
        {
            (10, 0.803275)
            (100, 0.818560)
            (1000, 0.846249)
            (10000, 0.887971)
            (100000, 0.939181)
            (1000000, 0.975705)
        };

        \legend{Recall, Precision, F-measure}

    \end{semilogxaxis}

\end{tikzpicture}

%% file: resultsROCSimple.tex
\begin{tikzpicture}

    \begin{axis}[legend cell align={left}, xmin=0.0, xmax=1.0, ymin=0.0, ymax=1.0, xlabel=False positives, ylabel=True positives, legend pos=south east]

        \addplot[draw] plot[mark=none, line width=1pt, color=cyan]
        coordinates
        {
            (3.349041e-04, 0.029081)
            (3.349041e-04, 0.034858)
            (1.221873e-03, 0.063729)
            (1.398314e-03, 0.069115)
            (2.597799e-03, 0.097108)
            (2.697183e-03, 0.103185)
            (3.764713e-03, 0.131798)
            (4.040105e-03, 0.137222)
            (5.586237e-03, 0.165031)
            (5.909039e-03, 0.170922)
            (7.360102e-03, 0.198999)
            (7.624286e-03, 0.204849)
            (8.797228e-03, 0.232903)
            (8.957616e-03, 0.238895)
            (9.948213e-03, 0.266877)
            (1.005585e-02, 0.273055)
            (1.106163e-02, 0.301640)
            (1.127669e-02, 0.307142)
            (1.255390e-02, 0.335216)
            (1.268134e-02, 0.341155)
            (1.388684e-02, 0.369621)
            (1.415679e-02, 0.374958)
            (1.567418e-02, 0.403057)
            (1.589540e-02, 0.408750)
            (1.767015e-02, 0.436787)
            (1.782009e-02, 0.442275)
            (1.930508e-02, 0.470069)
            (1.962882e-02, 0.475944)
            (2.079636e-02, 0.504104)
            (2.115166e-02, 0.509677)
            (2.282115e-02, 0.537221)
            (2.292456e-02, 0.543368)
            (2.453218e-02, 0.570949)
            (2.485390e-02, 0.576946)
            (2.655283e-02, 0.603956)
            (2.687167e-02, 0.610229)
            (2.847295e-02, 0.637060)
            (2.867607e-02, 0.643803)
            (3.059387e-02, 0.670172)
            (3.103225e-02, 0.676636)
            (3.300640e-02, 0.702394)
            (3.362786e-02, 0.708929)
            (3.652581e-02, 0.733666)
            (3.732523e-02, 0.740153)
            (4.067809e-02, 0.764612)
            (4.195354e-02, 0.770279)
            (4.583094e-02, 0.792904)
            (4.751364e-02, 0.799104)
            (5.290569e-02, 0.820295)
            (5.529244e-02, 0.825822)
            (6.200015e-02, 0.845521)
            (6.505959e-02, 0.850363)
            (7.339348e-02, 0.867876)
            (7.760196e-02, 0.872419)
            (8.771292e-02, 0.888158)
            (9.330834e-02, 0.891927)
            (1.043218e-01, 0.905928)
            (1.115791e-01, 0.909433)
            (1.250877e-01, 0.921554)
            (1.328622e-01, 0.924776)
            (1.482247e-01, 0.935053)
            (1.577084e-01, 0.937786)
            (1.742871e-01, 0.946573)
            (1.852095e-01, 0.949104)
            (2.041396e-01, 0.956084)
            (2.167039e-01, 0.958033)
            (2.373011e-01, 0.964218)
            (2.513075e-01, 0.965720)
            (2.727339e-01, 0.970637)
            (2.886368e-01, 0.971950)
            (3.123862e-01, 0.975606)
            (3.293315e-01, 0.976681)
            (3.530153e-01, 0.979763)
            (3.707755e-01, 0.980712)
            (3.953594e-01, 0.983449)
            (4.145580e-01, 0.984263)
            (4.399617e-01, 0.986353)
            (4.597875e-01, 0.987090)
            (4.853635e-01, 0.988607)
            (5.059300e-01, 0.989231)
            (5.323154e-01, 0.990688)
            (5.534333e-01, 0.991334)
            (5.789326e-01, 0.992595)
            (6.007047e-01, 0.993309)
            (6.265761e-01, 0.994442)
            (6.491234e-01, 0.995038)
            (6.743178e-01, 0.996105)
            (6.974536e-01, 0.996852)
            (7.224689e-01, 0.997289)
            (7.472213e-01, 0.997630)
            (7.719272e-01, 0.998060)
            (7.974564e-01, 0.998399)
            (8.225653e-01, 0.998746)
            (8.483082e-01, 0.999069)
            (8.731018e-01, 0.999388)
            (8.982309e-01, 0.999639)
            (9.233955e-01, 0.999880)
            (9.488935e-01, 0.999983)
            (9.743093e-01, 1.000000)
            (1.000000e+00, 1.000000)
        };

        \addplot[draw] plot[mark=none, line width=1pt, color=purple]
        coordinates
        {
            (4.616317e-04, 0.028650)
            (5.413390e-04, 0.034586)
            (1.933072e-03, 0.062646)
            (2.293878e-03, 0.068195)
            (4.358287e-03, 0.095875)
            (4.689296e-03, 0.101270)
            (6.924115e-03, 0.128681)
            (7.357365e-03, 0.134117)
            (9.602109e-03, 0.161570)
            (1.000006e-02, 0.166911)
            (1.241495e-02, 0.194148)
            (1.287227e-02, 0.199486)
            (1.537554e-02, 0.226828)
            (1.582408e-02, 0.232203)
            (1.832414e-02, 0.260137)
            (1.867494e-02, 0.265004)
            (2.134693e-02, 0.292527)
            (2.178957e-02, 0.297555)
            (2.457223e-02, 0.325362)
            (2.494684e-02, 0.330004)
            (2.792051e-02, 0.357192)
            (2.847286e-02, 0.362250)
            (3.206966e-02, 0.389339)
            (3.264884e-02, 0.394047)
            (3.603864e-02, 0.420770)
            (3.657918e-02, 0.425907)
            (3.999886e-02, 0.451983)
            (4.072425e-02, 0.457388)
            (4.346947e-02, 0.484597)
            (4.404687e-02, 0.489564)
            (4.723605e-02, 0.516621)
            (4.783971e-02, 0.521478)
            (5.117372e-02, 0.547481)
            (5.164036e-02, 0.553355)
            (5.515773e-02, 0.579065)
            (5.586650e-02, 0.585061)
            (5.982728e-02, 0.610098)
            (6.076632e-02, 0.616141)
            (6.488089e-02, 0.640793)
            (6.565057e-02, 0.646965)
            (7.043204e-02, 0.670798)
            (7.146268e-02, 0.676908)
            (7.659999e-02, 0.700790)
            (7.781352e-02, 0.706409)
            (8.345701e-02, 0.729189)
            (8.504105e-02, 0.734545)
            (9.116686e-02, 0.756654)
            (9.281287e-02, 0.761821)
            (1.004565e-01, 0.782171)
            (1.026100e-01, 0.787481)
            (1.114316e-01, 0.806511)
            (1.141260e-01, 0.811430)
            (1.252604e-01, 0.828016)
            (1.290062e-01, 0.832749)
            (1.412271e-01, 0.848475)
            (1.454571e-01, 0.852428)
            (1.590343e-01, 0.867104)
            (1.641034e-01, 0.870573)
            (1.789914e-01, 0.883597)
            (1.851710e-01, 0.887074)
            (2.018241e-01, 0.898174)
            (2.095359e-01, 0.901232)
            (2.277224e-01, 0.911347)
            (2.368383e-01, 0.913834)
            (2.575809e-01, 0.921944)
            (2.676648e-01, 0.924304)
            (2.883273e-01, 0.931746)
            (2.997038e-01, 0.933949)
            (3.217514e-01, 0.940848)
            (3.340421e-01, 0.942957)
            (3.571842e-01, 0.948370)
            (3.713378e-01, 0.950032)
            (3.940071e-01, 0.955517)
            (4.086760e-01, 0.957123)
            (4.321573e-01, 0.962138)
            (4.476186e-01, 0.963726)
            (4.725019e-01, 0.967800)
            (4.889003e-01, 0.969363)
            (5.135638e-01, 0.972932)
            (5.307196e-01, 0.974640)
            (5.548740e-01, 0.978186)
            (5.727913e-01, 0.979707)
            (5.962122e-01, 0.982931)
            (6.153421e-01, 0.984599)
            (6.394859e-01, 0.987073)
            (6.601162e-01, 0.988637)
            (6.837895e-01, 0.990504)
            (7.053554e-01, 0.991923)
            (7.291862e-01, 0.993288)
            (7.525915e-01, 0.994264)
            (7.767046e-01, 0.995263)
            (8.006678e-01, 0.996221)
            (8.257826e-01, 0.997016)
            (8.504463e-01, 0.997540)
            (8.748105e-01, 0.998360)
            (8.992706e-01, 0.998968)
            (9.242448e-01, 0.999444)
            (9.491318e-01, 0.999733)
            (9.744511e-01, 0.999937)
            (1.000000e+00, 1.000000)
        };

        \addplot[draw] plot[mark=none, line width=1pt, color=orange]
        coordinates
        {
            (1.259298e-03, 0.028299)
            (1.416833e-03, 0.033919)
            (4.232988e-03, 0.061276)
            (4.661004e-03, 0.066286)
            (7.767496e-03, 0.093360)
            (8.290404e-03, 0.098234)
            (1.185477e-02, 0.124945)
            (1.243066e-02, 0.129731)
            (1.658268e-02, 0.155634)
            (1.733039e-02, 0.160711)
            (2.217623e-02, 0.186499)
            (2.303877e-02, 0.191365)
            (2.722590e-02, 0.217452)
            (2.817385e-02, 0.222374)
            (3.291240e-02, 0.248484)
            (3.392208e-02, 0.252670)
            (3.916836e-02, 0.278464)
            (3.996520e-02, 0.282842)
            (4.493254e-02, 0.308339)
            (4.592285e-02, 0.312918)
            (5.204484e-02, 0.337472)
            (5.295455e-02, 0.342196)
            (5.949992e-02, 0.366619)
            (6.060333e-02, 0.370769)
            (6.678486e-02, 0.395527)
            (6.798962e-02, 0.399762)
            (7.435277e-02, 0.424585)
            (7.533147e-02, 0.428819)
            (8.082993e-02, 0.453729)
            (8.217604e-02, 0.458331)
            (8.882260e-02, 0.482197)
            (9.031972e-02, 0.486717)
            (9.721584e-02, 0.510726)
            (9.836159e-02, 0.515154)
            (1.054267e-01, 0.537933)
            (1.071320e-01, 0.543105)
            (1.141858e-01, 0.565754)
            (1.160440e-01, 0.570960)
            (1.234702e-01, 0.593619)
            (1.250096e-01, 0.598469)
            (1.327360e-01, 0.621254)
            (1.343447e-01, 0.625933)
            (1.425453e-01, 0.647544)
            (1.442483e-01, 0.652534)
            (1.533722e-01, 0.673497)
            (1.552878e-01, 0.678171)
            (1.646031e-01, 0.698406)
            (1.664444e-01, 0.703323)
            (1.772304e-01, 0.722386)
            (1.798685e-01, 0.727043)
            (1.898334e-01, 0.745771)
            (1.927868e-01, 0.750369)
            (2.053838e-01, 0.767545)
            (2.088990e-01, 0.771482)
            (2.220523e-01, 0.787286)
            (2.262494e-01, 0.791413)
            (2.399858e-01, 0.807030)
            (2.445936e-01, 0.810948)
            (2.600183e-01, 0.824961)
            (2.654831e-01, 0.828758)
            (2.832084e-01, 0.840705)
            (2.897358e-01, 0.844214)
            (3.074813e-01, 0.855717)
            (3.143239e-01, 0.859401)
            (3.333136e-01, 0.870412)
            (3.409515e-01, 0.873570)
            (3.604770e-01, 0.884006)
            (3.687356e-01, 0.887421)
            (3.890211e-01, 0.896486)
            (3.983783e-01, 0.899747)
            (4.190851e-01, 0.908342)
            (4.293690e-01, 0.911665)
            (4.506842e-01, 0.919386)
            (4.620083e-01, 0.922345)
            (4.832171e-01, 0.929647)
            (4.953039e-01, 0.932867)
            (5.167941e-01, 0.939301)
            (5.300724e-01, 0.942714)
            (5.514048e-01, 0.949155)
            (5.653875e-01, 0.952274)
            (5.876548e-01, 0.958018)
            (6.026987e-01, 0.961116)
            (6.239090e-01, 0.966249)
            (6.400136e-01, 0.969570)
            (6.612158e-01, 0.973938)
            (6.787574e-01, 0.976999)
            (6.997265e-01, 0.980606)
            (7.191408e-01, 0.983612)
            (7.396278e-01, 0.986520)
            (7.612942e-01, 0.988746)
            (7.837658e-01, 0.990889)
            (8.067941e-01, 0.992718)
            (8.302772e-01, 0.994247)
            (8.537918e-01, 0.995549)
            (8.776570e-01, 0.996594)
            (9.014902e-01, 0.997694)
            (9.251206e-01, 0.998932)
            (9.495343e-01, 0.999604)
            (9.748427e-01, 0.999877)
            (1.000000e+00, 1.000000)
        };

        \addplot[draw] plot[mark=none, line width=1pt, color=green]
        coordinates
        {
            (1.354277e-03, 0.028418)
            (1.737403e-03, 0.033706)
            (5.036964e-03, 0.060322)
            (5.773141e-03, 0.065258)
            (9.481877e-03, 0.091438)
            (1.005365e-02, 0.096621)
            (1.497434e-02, 0.122667)
            (1.574457e-02, 0.126922)
            (2.088440e-02, 0.152609)
            (2.175136e-02, 0.157218)
            (2.681005e-02, 0.182398)
            (2.789528e-02, 0.187215)
            (3.388561e-02, 0.212197)
            (3.509710e-02, 0.216222)
            (4.125281e-02, 0.240850)
            (4.234059e-02, 0.245300)
            (4.872061e-02, 0.270441)
            (4.957193e-02, 0.274682)
            (5.722118e-02, 0.297889)
            (5.839224e-02, 0.302227)
            (6.509895e-02, 0.326146)
            (6.637422e-02, 0.330335)
            (7.541087e-02, 0.353313)
            (7.650302e-02, 0.357107)
            (8.404375e-02, 0.380631)
            (8.550454e-02, 0.384590)
            (9.400442e-02, 0.407535)
            (9.566803e-02, 0.411420)
            (1.048622e-01, 0.433433)
            (1.062083e-01, 0.437910)
            (1.152209e-01, 0.459671)
            (1.174056e-01, 0.463496)
            (1.271854e-01, 0.484802)
            (1.291058e-01, 0.488904)
            (1.390681e-01, 0.509435)
            (1.416456e-01, 0.514073)
            (1.520485e-01, 0.534418)
            (1.543852e-01, 0.538580)
            (1.649112e-01, 0.558370)
            (1.672139e-01, 0.563215)
            (1.778425e-01, 0.582905)
            (1.800544e-01, 0.587658)
            (1.907516e-01, 0.607156)
            (1.934690e-01, 0.611501)
            (2.058989e-01, 0.629751)
            (2.086723e-01, 0.634280)
            (2.201497e-01, 0.653306)
            (2.226912e-01, 0.657917)
            (2.356937e-01, 0.675114)
            (2.393084e-01, 0.679275)
            (2.536710e-01, 0.695689)
            (2.574237e-01, 0.699982)
            (2.713660e-01, 0.716581)
            (2.748460e-01, 0.721310)
            (2.907359e-01, 0.735987)
            (2.951874e-01, 0.740477)
            (3.101720e-01, 0.755487)
            (3.148556e-01, 0.759734)
            (3.304146e-01, 0.774444)
            (3.356643e-01, 0.778373)
            (3.533036e-01, 0.791747)
            (3.582227e-01, 0.795489)
            (3.745080e-01, 0.809183)
            (3.800212e-01, 0.813507)
            (3.977837e-01, 0.825653)
            (4.040717e-01, 0.830071)
            (4.221489e-01, 0.842015)
            (4.290042e-01, 0.845901)
            (4.479238e-01, 0.857000)
            (4.554940e-01, 0.861167)
            (4.745085e-01, 0.871336)
            (4.829137e-01, 0.875833)
            (5.013771e-01, 0.885501)
            (5.110128e-01, 0.889995)
            (5.296980e-01, 0.899595)
            (5.399658e-01, 0.903654)
            (5.580484e-01, 0.912628)
            (5.687362e-01, 0.917858)
            (5.879527e-01, 0.925647)
            (5.998187e-01, 0.930690)
            (6.185989e-01, 0.937810)
            (6.316865e-01, 0.942630)
            (6.496800e-01, 0.950217)
            (6.629816e-01, 0.955088)
            (6.806555e-01, 0.961581)
            (6.956034e-01, 0.966658)
            (7.130312e-01, 0.972640)
            (7.294586e-01, 0.977055)
            (7.489764e-01, 0.980922)
            (7.686379e-01, 0.984094)
            (7.898302e-01, 0.986480)
            (8.121966e-01, 0.988822)
            (8.348125e-01, 0.990773)
            (8.575618e-01, 0.993012)
            (8.799697e-01, 0.995002)
            (9.031922e-01, 0.996643)
            (9.264340e-01, 0.998086)
            (9.504764e-01, 0.999148)
            (9.747571e-01, 0.999872)
            (1.000000e+00, 1.000000)
        };

        \addplot[draw] plot[mark=none, line width=1pt, color=blue]
        coordinates
        {
            (3.298373e-03, 0.027108)
            (3.881789e-03, 0.031929)
            (8.788664e-03, 0.056939)
            (9.898341e-03, 0.061927)
            (1.561580e-02, 0.086277)
            (1.691242e-02, 0.091055)
            (2.318673e-02, 0.114795)
            (2.451928e-02, 0.119840)
            (3.056199e-02, 0.143763)
            (3.172932e-02, 0.148690)
            (3.828217e-02, 0.171996)
            (3.976425e-02, 0.176985)
            (4.690055e-02, 0.200180)
            (4.866776e-02, 0.204736)
            (5.610516e-02, 0.227847)
            (5.782333e-02, 0.232452)
            (6.569267e-02, 0.255008)
            (6.788073e-02, 0.259526)
            (7.641471e-02, 0.281695)
            (7.811188e-02, 0.285871)
            (8.723931e-02, 0.307997)
            (8.868854e-02, 0.312200)
            (9.735184e-02, 0.333634)
            (9.951915e-02, 0.338259)
            (1.088187e-01, 0.359232)
            (1.107799e-01, 0.363964)
            (1.203407e-01, 0.384429)
            (1.228457e-01, 0.389322)
            (1.329155e-01, 0.409342)
            (1.358175e-01, 0.413766)
            (1.461347e-01, 0.433427)
            (1.493618e-01, 0.438149)
            (1.601147e-01, 0.456890)
            (1.628246e-01, 0.462238)
            (1.733810e-01, 0.481281)
            (1.764565e-01, 0.486711)
            (1.865409e-01, 0.505637)
            (1.896791e-01, 0.511160)
            (2.011614e-01, 0.529222)
            (2.044702e-01, 0.534773)
            (2.157060e-01, 0.552135)
            (2.195138e-01, 0.558085)
            (2.313955e-01, 0.574449)
            (2.352382e-01, 0.580698)
            (2.479716e-01, 0.596968)
            (2.520015e-01, 0.602551)
            (2.642822e-01, 0.618258)
            (2.684504e-01, 0.624156)
            (2.812870e-01, 0.639705)
            (2.859378e-01, 0.645488)
            (2.987579e-01, 0.660762)
            (3.031461e-01, 0.666695)
            (3.171678e-01, 0.681211)
            (3.223598e-01, 0.687113)
            (3.373794e-01, 0.701195)
            (3.427919e-01, 0.706603)
            (3.570165e-01, 0.720525)
            (3.625893e-01, 0.726352)
            (3.769675e-01, 0.739504)
            (3.832452e-01, 0.745673)
            (3.974825e-01, 0.758973)
            (4.042169e-01, 0.764443)
            (4.188404e-01, 0.777251)
            (4.258172e-01, 0.783421)
            (4.414556e-01, 0.794650)
            (4.495071e-01, 0.800596)
            (4.645497e-01, 0.812225)
            (4.724958e-01, 0.817935)
            (4.874312e-01, 0.829572)
            (4.950798e-01, 0.836080)
            (5.111804e-01, 0.846665)
            (5.202584e-01, 0.852798)
            (5.354639e-01, 0.862223)
            (5.456175e-01, 0.868647)
            (5.594792e-01, 0.879117)
            (5.703862e-01, 0.885208)
            (5.851026e-01, 0.894827)
            (5.963835e-01, 0.901078)
            (6.104732e-01, 0.910570)
            (6.214485e-01, 0.917353)
            (6.372155e-01, 0.925189)
            (6.498013e-01, 0.931381)
            (6.655169e-01, 0.939263)
            (6.784897e-01, 0.945575)
            (6.943366e-01, 0.952381)
            (7.093756e-01, 0.957947)
            (7.256335e-01, 0.964053)
            (7.407936e-01, 0.969765)
            (7.579070e-01, 0.975205)
            (7.759871e-01, 0.979252)
            (7.963513e-01, 0.982236)
            (8.175416e-01, 0.985384)
            (8.389499e-01, 0.988195)
            (8.607021e-01, 0.991021)
            (8.826173e-01, 0.993514)
            (9.046134e-01, 0.995853)
            (9.269012e-01, 0.997647)
            (9.503663e-01, 0.999055)
            (9.747652e-01, 0.999862)
            (1.000000e+00, 1.000000)
        };

        \addplot[draw] plot[mark=none, line width=1pt, color=red]
        coordinates
        {
            (6.274810e-03, 0.023139)
            (8.580286e-03, 0.028248)
            (1.524521e-02, 0.050593)
            (1.773354e-02, 0.055835)
            (2.484515e-02, 0.077299)
            (2.745444e-02, 0.083091)
            (3.455838e-02, 0.104385)
            (3.788798e-02, 0.109733)
            (4.584215e-02, 0.129703)
            (4.910766e-02, 0.135307)
            (5.636154e-02, 0.156348)
            (5.893726e-02, 0.162728)
            (6.652017e-02, 0.183089)
            (6.871251e-02, 0.189648)
            (7.643521e-02, 0.209780)
            (7.904973e-02, 0.216273)
            (8.718742e-02, 0.236591)
            (8.973643e-02, 0.243299)
            (9.791667e-02, 0.262663)
            (1.008828e-01, 0.269280)
            (1.096130e-01, 0.287919)
            (1.126747e-01, 0.294860)
            (1.217770e-01, 0.313354)
            (1.254871e-01, 0.319652)
            (1.351599e-01, 0.337565)
            (1.387689e-01, 0.344297)
            (1.475684e-01, 0.362084)
            (1.508220e-01, 0.369150)
            (1.594902e-01, 0.386969)
            (1.633849e-01, 0.394320)
            (1.725321e-01, 0.411426)
            (1.770252e-01, 0.418086)
            (1.874439e-01, 0.434023)
            (1.920598e-01, 0.441355)
            (2.005937e-01, 0.457642)
            (2.052747e-01, 0.465725)
            (2.145494e-01, 0.481790)
            (2.196010e-01, 0.489474)
            (2.293217e-01, 0.505160)
            (2.341824e-01, 0.512914)
            (2.453496e-01, 0.527744)
            (2.512842e-01, 0.535318)
            (2.617272e-01, 0.549714)
            (2.672718e-01, 0.557874)
            (2.785538e-01, 0.571712)
            (2.847821e-01, 0.579484)
            (2.963202e-01, 0.592283)
            (3.026658e-01, 0.600335)
            (3.133616e-01, 0.614627)
            (3.196498e-01, 0.621935)
            (3.311519e-01, 0.635087)
            (3.376001e-01, 0.643875)
            (3.480766e-01, 0.656934)
            (3.547158e-01, 0.665168)
            (3.660858e-01, 0.677158)
            (3.738486e-01, 0.685662)
            (3.852272e-01, 0.697751)
            (3.927488e-01, 0.706312)
            (4.036104e-01, 0.718489)
            (4.118064e-01, 0.726661)
            (4.224975e-01, 0.738750)
            (4.306241e-01, 0.746799)
            (4.426162e-01, 0.758189)
            (4.513239e-01, 0.766792)
            (4.638313e-01, 0.777354)
            (4.725384e-01, 0.785297)
            (4.854949e-01, 0.795360)
            (4.948069e-01, 0.803762)
            (5.069550e-01, 0.813937)
            (5.165715e-01, 0.822033)
            (5.281218e-01, 0.832652)
            (5.383589e-01, 0.840648)
            (5.512506e-01, 0.849914)
            (5.627986e-01, 0.857455)
            (5.760490e-01, 0.866594)
            (5.875562e-01, 0.874245)
            (6.007272e-01, 0.883184)
            (6.124622e-01, 0.891176)
            (6.258743e-01, 0.899602)
            (6.383429e-01, 0.907048)
            (6.519838e-01, 0.915546)
            (6.640102e-01, 0.923184)
            (6.779687e-01, 0.930485)
            (6.920719e-01, 0.937683)
            (7.061169e-01, 0.944524)
            (7.202117e-01, 0.951358)
            (7.352527e-01, 0.958078)
            (7.499768e-01, 0.964251)
            (7.675954e-01, 0.969190)
            (7.863399e-01, 0.972798)
            (8.068081e-01, 0.976306)
            (8.274991e-01, 0.979428)
            (8.481121e-01, 0.982869)
            (8.682484e-01, 0.986331)
            (8.881856e-01, 0.989825)
            (9.086535e-01, 0.993190)
            (9.301514e-01, 0.996021)
            (9.522640e-01, 0.998106)
            (9.753885e-01, 0.999501)
            (1.000000e+00, 1.000000)
        };

        \draw[black, dotted] (axis cs:\pgfkeysvalueof{/pgfplots/xmin},\pgfkeysvalueof{/pgfplots/xmin}) -- (axis cs:\pgfkeysvalueof{/pgfplots/xmax},\pgfkeysvalueof{/pgfplots/xmax});

        \legend{$\M=10^6$, $\M=10^5$, $\M=10^4$, $\M=10^3$, $\M=10^2$, $\M=10$}

    \end{axis}

\end{tikzpicture}

%% file: fmeasureVaryingN.tex
\begin{tikzpicture}

    \begin{axis}[legend cell align={left}, xmin=10, xmax=100, ymin=0.0, ymax=1.0, xlabel=$N$, ylabel=F-measure, legend pos=north east]

        \addplot plot[mark=none, line width=1pt, color=green]
        coordinates
        {
            (10, 0.938016)
            (20, 0.742337)
            (30, 0.570442)
            (40, 0.433884)
            (50, 0.356892)
            (60, 0.307914)
            (70, 0.264899)
            (80, 0.232234)
            (90, 0.207437)
            (100, 0.189095)
        };

        \addplot plot[mark=none, line width=1pt, color=green, dashed]
        coordinates
        {
            (10, 0.766401)
            (20, 0.554704)
            (30, 0.450093)
            (40, 0.382110)
            (50, 0.335427)
            (60, 0.297527)
            (70, 0.266190)
            (80, 0.247224)
            (90, 0.226441)
            (100, 0.210500)
        };

        \addplot plot[mark=none, line width=1pt, color=red]
        coordinates
        {
            (10, 0.893554)
            (20, 0.606167)
            (30, 0.433729)
            (40, 0.334686)
            (50, 0.272711)
            (60, 0.231027)
            (70, 0.202455)
            (80, 0.181600)
            (90, 0.168808)
            (100, 0.153841)
        };

        \addplot plot[mark=none, line width=1pt, color=red, dashed]
        coordinates
        {
            (10, 0.908429)
            (20, 0.650244)
            (30, 0.491652)
            (40, 0.406569)
            (50, 0.369390)
            (60, 0.318308)
            (70, 0.283808)
            (80, 0.268094)
            (90, 0.250662)
            (100, 0.237769)
        };

        \addplot plot[mark=none, line width=1pt, color=blue]
        coordinates
        {
            (10, 0.363692)
            (20, 0.225801)
            (30, 0.143626)
            (40, 0.115698)
            (50, 0.089433)
            (60, 0.076058)
            (70, 0.065135)
            (80, 0.056790)
            (90, 0.050325)
            (100, 0.045945)
        };

        \addplot plot[mark=none, line width=1pt, color=blue, dashed]
        coordinates
        {
            (10, 0.487002)
            (20, 0.295346)
            (30, 0.210564)
            (40, 0.160651)
            (50, 0.126282)
            (60, 0.104029)
            (70, 0.086904)
            (80, 0.072934)
            (90, 0.064044)
            (100, 0.057055)
        };

        \legend{\emph{Simple} (RG), \emph{Sparse} (RG), \emph{Simple} (ER), \emph{Sparse} (ER), \emph{Simple} (ring), \emph{Sparse} (ring)}

    \end{axis}

\end{tikzpicture}

%% file: graphSignalsAdequation.tex
\begin{tikzpicture}

    \begin{axis}[legend cell align={left}, width=8.0cm, height=4.0cm, xmin=10, xmax=200, ymin=0.4, ymax=1.0, xlabel=$M$, ylabel=Success ratio, legend pos=south east]

        \addplot plot[mark=none, line width=1pt, color=red]
        coordinates
        {
            (10, 0.49)
            (20, 0.5975)
            (30, 0.665)
            (40, 0.7175)
            (50, 0.75)
            (60, 0.765)
            (70, 0.78)
            (80, 0.8)
            (90, 0.8175)
            (100, 0.82)
            (110, 0.8375)
            (120, 0.8575)
            (130, 0.87)
            (140, 0.87)
            (150, 0.8825)
            (160, 0.885)
            (170, 0.8925)
            (180, 0.8975)
            (190, 0.9025)
            (200, 0.915)
        };

    \end{axis}

\end{tikzpicture}

%% file: resultsBothHistograms.tex
\begin{tikzpicture}
    \begin{axis}[legend cell align={left}, ybar, width=8.0cm, height=4.2cm, ymin=0, xlabel=Distance to polytope, ylabel=Occurrences]
        \addplot +[hist={bins=212, data min=0.0, data max=2.112693}] table [y index=0] {resultsKalofoliasHistogramData.csv};
        \addplot +[hist={bins=212, data min=0.0, data max=2.112693}, opacity=0.5] table [y index=0] {resultsRandomHistogramData.csv};
        \legend{Kalofolias (${\bf d}_K$),Random (${\bf d}_R$)}
    \end{axis}
\end{tikzpicture}

%% file: acknowledgements.tex
    \section*{Acknowledgements}
    \label{acknowledgements}
        
        The authors would like to thank the reviewers of previous versions of this paper, whose remarks helped improving the quality of our work.
        Also, we would like to thank Xiaowen Dong and Santiago Segarra for kindly providing their codes, as well as Benjamin Girault for sharing his dataset.
        Additionally, we would like to thank Pierre Vandergheynst and his group at EPFL for the inspiring discussions that led to this work.
        
    %
